\documentclass[]{aa} 
 
\usepackage{graphicx, times, float, color, hyperref} 
\usepackage[figuresright]{rotating}
\usepackage[varg]{txfonts}

\newcommand{\less}{\raisebox{-1.1mm}{$\stackrel{<}{\sim}$}} 
\newcommand{\more}{\raisebox{-1.1mm}{$\stackrel{>}{\sim}$}} 
\newcommand{\msol}{\mbox{$M_{\odot}$}} 
 
\newcommand{\rstar}{\mbox{$R_{\star}$}} 
\newcommand{\msolyr}{{$M_{\odot}$}\,yr$^{-1}$} 
\newcommand{\mdot}{$\dot{M}$}
\newcommand{\lsol}{\mbox{$L_{\odot}$}}

\newcommand{\ks}{km s$^{-1}$} 
 
\newcommand{\mum}{$\mu$m}

\begin{document}

\title{
Luminosities and mass-loss rates of Local Group AGB stars and Red Supergiants\thanks{
Tables~\ref{Tab-Csample}, \ref{Tab-Msample}, \ref{Tab-Cstar}, \ref{Tab-Cstar}, and \ref{Tab-Synphot} are available in 
electronic form at the CDS via anonymous ftp to cdsarc.u-strasbg.fr (130.79.128.5) or via 
http://cdsweb.u-strasbg.fr/cgi-bin/qcat?J/A+A/. 
Figures~\ref{Fig-Cstars} and \ref{Fig-Mstars} are available in the on-line edition of A\&A. 
}
} 
 
\author{M.~A.~T.~Groenewegen \inst{1}  
\and 
G.~C.~Sloan \inst{2,3,4} 
}

\institute{ 
Koninklijke Sterrenwacht van Belgi\"e, Ringlaan 3, B--1180 Brussels, Belgium \\ \email{martin.groenewegen@oma.be}
\and
Cornell Center for Astrophysics and Planetary Science, Cornell University, 
Ithaca, NY 14853-6801, USA
\and
Department of Physics and Astronomy, University of North Carolina, 
Chapel Hill, NC 27599-3255, USA
\and
Space Telescope Science Institute, 3700 San Martin Dr., 
Baltimore, MD 21218, USA
}

\date{received: A\&A submitted 2 May 2017,  accepted: 2017} 
 
\offprints{Martin Groenewegen} 
 
 
\abstract {
Mass loss is one of the fundamental properties of asymptotic 
giant branch (AGB) stars, and through the enrichment of the 
interstellar medium, AGB stars are key players in the life 
cycle of dust and gas in the universe.  However, a 
quantitative understanding of the mass-loss process is still 
largely lacking.
}
{We aim to investigate mass loss and luminosity in a 
large sample of evolved stars  in several Local Group 
galaxies with a variety of metalliticies and star-formation 
histories: the Small and Large Magellanic Cloud, and the 
Fornax, Carina, and Sculptor dwarf spheroidal galaxies (dSphs).
}
{Dust radiative transfer models are presented for 225 carbon 
stars and 171 oxygen-rich evolved stars in several
Local Group galaxies for which spectra from
the Infrared Spectrograph on {\it Spitzer} are 
available.  The spectra are complemented with available 
optical and infrared photometry to construct spectral 
energy distributions.  A minimization procedure was used to 
determine luminosity and mass-loss rate (MLR).  Pulsation 
periods were derived for a large fraction of the sample based 
on a re-analysis of existing data.
}
{
New deep $K$-band photometry from the VMC survey and 
multi-epoch data from IRAC (at 4.5~\mum) and 
AllWISE  and NEOWISE have allowed us to derive pulsation 
periods longer than 1000 days for some of the most heavily 
obscured and reddened objects.
We derive (dust) MLRs and luminosities for the entire sample. 
The estimated MLRs can differ significantly from estimates
for the same objects in the literature due to differences
in adopted optical constants (up to factors of several) 
and details in the radiative transfer modelling.
Updated parameters for the super-AGB candidate MSX SMC 055 
(IRAS 00483$-$7347) are presented.  Its current mass is 
estimated to be 8.5 $\pm$ 1.6~\msol, suggesting an initial 
mass well above 8~\msol\ in agreement with 
estimates based on its large Rubidium abundance.
Using synthetic photometry, we present and discuss
colour-colour and colour-magnitude diagrams which can be
expected from the James Webb Space Telescope.
}
{}


\maketitle 
 
\section{Introduction} 

Almost all stars with initial masses in the range $\sim$ 
0.9--8~\msol\ will pass through the asymptotic giant branch
(AGB) phase, which is the final stage of active nuclear 
burning before they become post-AGB objects, planetary 
nebulae and finally white dwarfs.  Stars above this mass 
range will pass through the red supergiant (RSG) phase 
before they may end as supernovae.  In both cases, mass loss 
dominates the final evolutionary stages of the star.

The {\it Infrared Astronomical Satellite (IRAS)} and the 
{\it Infrared Space Observatory (ISO)} have greatly 
improved our understanding of these final evolutionary 
stages for stars in the Galaxy, but uncertainties in 
distances prevent accurate luminosities and mass-loss rates 
(MLRs).  Sources at known distances, as in the Large and 
Small Magellanic Clouds (LMC and SMC), or nearby dwarf 
spheroidal galaxies (dSphs), reduce this problem and also 
enable the study of the effect of metallicity on the MLR.
Surveys of the Magellanic Clouds (MCs) with the {\it Spitzer Space Telescope} 
(Werner et al.\ 2004) have added to this legacy with data
from IRAC (Fazio et al.\ 2004) and MIPS (Rieke et al.\ 2004):
the {\it Spitzer} Survey of the SMC (S$^3$MC; Bolatto et al.\ 
2007) and the full-galaxy catalogues from {\it Surveying the 
Agents of Galactic Evolution} program for the LMC (Meixner 
et al.\ 2006) and SMC (Gordon et al.\ 2011).

Groenewegen et al.\ (2007) previously modelled the spectral 
energy distributions (SEDs) and spectra taken with the 
Infrared Spectrograph (IRS; Houck et al.\ 2004) on board 
{\it Spitzer} for a 
sample of 60 carbon (C) stars.  Groenewegen et al.\ (2009, 
hereafter G09) extended this to 101 C stars and 86 
oxygen-rich AGB stars and RSGs (hereafter referred to as M 
stars for simplicity) in the MCs.

The cryogenic phase of the {\it Spitzer} mission is over and 
the data taken with the {\it IRS} spectrograph are publicly 
available, for example through the CASSIS website (Lebouteiller et 
al.\ 2011)\footnote{Combined Atlas of {\it Spitzer}/IRS 
sources, available at http://cassis.sirtf.com.}.  In 
this paper we investigate a sample of AGB stars and RSGs in 
the MCs that is almost double in size compared to that 
considered by G09, including 19 C stars in four dSphs: 
Fornax, Carina, Leo I, and Sculptor (Matsuura et al.\ 2007, 
Sloan et al.\ 2012).

Considerable work is ongoing to model large numbers of AGB 
stars in the MCs using the available photometry, by 
fitting the SEDs of individual stars with a radiative transfer 
model.  Gullieuszik et al.\ (2012) presented results on 374 
AGB candidates in a 1.42 square degree area\footnote{An area 
corresponding to a single `tile' observed by VISTA, out of a 
planned 180 square degrees for the final VMC survey covering 
SMC, LMC, Bridge and Stream.} in the LMC observed as part of 
the VISTA Magellanic Cloud Survey (VMC; Cioni et al. 2011).  
Riebel et al.\ (2012) derived MLRs for a sample of 
$\sim$30~000 AGB stars and RSGs in the LMC, by fitting up to 
12 bands of photometry to the precomputed Grid of Red 
Supergiant and Asymptotic Giant Branch ModelS (GRAMS; 
Srinivasan et al.\ 2011 for the C-rich grid; Sargent et al.\ 
2011 for the O-rich grid). 
Srinivasan et al.\ (2016) used a similar approach for the SMC. 
Boyer et al.\ (2012) took a 
hybrid approach by first determining the dust MLR using the 
GRAMS models for 65 stars in five classes of AGB stars and RSGs.  
They then determined relations between the dust MLR and 
photometric excess at 8~\mum\ and used these to estimate the 
dust MLR for a total of about 25~000 AGB stars and RSGs in 
the LMC and about 7500 in the SMC.

To only consider stars with IRS spectra naturally limits the 
number of stars for which mass-loss rates and luminosities 
may be determined, but the spectra provide some 
important advantages compared to just a photometric sample.  
First, molecular absorption bands and dust emission features 
in the spectra allow confident identifications of C-rich 
versus O-rich stars (as described succinctly by Kraemer et 
al.\ 2002).  Spectroscopic classifications are not perfect,
primarily because some sources exhibit C-rich and O-rich
characteristics simultaneously, but they do help break the
degeneracies where the two classes overlap in colour-colour
space (for the bluest and reddest sources).  Second, the
spectra better constrain mineralogical properties of the
dust such as grain size and shape, crystallinity, and
chemistry.  And, because the spectra provide more information
on a source than the photometry can, they also better
constrain the radiative transfer models fitted to the data.
The lessons learned from the smaller spectroscopic sample can
then be applied to models of the much larger photometric samples.

Section~2 describes the sample of AGB stars and RSGs 
with IRS spectra and the spectral data.  Section~3 
describes the ancillary photometry and the periods 
derived from those data.  Section~4 presents the radiative 
transfer model and the properties of the dust species 
considered.  Section~5 presents the results and 
compares them to previous efforts (e.g.\ G09). 
In Sect.~6 we discuss the relation between stellar evolution and mass loss.

\section{The sample of infrared spectra} 

\subsection{Spectra} 

Several groups have obtained {\it Spitzer} IRS data of evolved 
stars in the LMC and SMC.  Table~\ref{Tab-Programs} lists the 
programs considered here. The last program listed provided 
data on a sample of 19 C stars in the Carina, Fornax, Leo I,
and Sculptor dSphs.

\begin{table*}
\setlength{\tabcolsep}{1.05mm}

\caption{Spectroscopic {\it Spitzer} programs considered in this paper.}
\begin{tabular}{rlllrr}
\hline
Program    & Principal              & Reference paper                                & Notes \\
  ID       & investigator           &                                                &        \\
\hline
  200      & J.~R.~Houck            & Sloan et al.\ (2008)                            & Evolved stars in the LMC and SMC \\
 1094      & F.~Kemper              &                                                 & AGB evolution in the LMC (and Galaxy) \\
 3277      & M.~P.~Egan             & Sloan et al.\ (2006), Kraemer et al.\ (2016)    & Infrared-bright sample in the SMC \\
 3426      & J.~H.~Kastner          & Buchanan et al.\ (2006)                         & Infrared-bright sample in the LMC \\
 3505      & P.~R.~Wood             & Zijlstra et al.\ (2006), Lagadec et al.\ (2007) & AGB stars in the LMC and SMC \\
 3591      & F.~Kemper              & Leisenring et al.\ (2008)                       & Evolved O-rich stars in the LMC \\
30155      & J.~R.~Houck            &                                                 & Sources with crystalline silicates in the SMC \\
30788      & R.~Sahai               & Sloan et al.\ (2014)                            & Embedded carbon stars and post-AGB objects in the LMC \\
40159      & A.~Tielens             & Kemper et al.\ (2010), Woods et al.\ (2011)     & Filling colour-colour and colour-mag.\ space in the LMC \\ 
40650      & L.~W.~Looney           & Gruendl et al.\ (2008)                          & YSOs (and some deeply embedded carbon stars) in the LMC \\
50167      & G.~Clayton             &                                                 & RSGs in the LMC and SMC \\
50240      & G.~C.~Sloan            &                                                 & Filling colour-colour and colour-mag.\ space in the SMC \\
50338      & M.~Matsuura            & Matsuura et al.\ (2014)                         & Carbon-rich post-AGB candidates in the LMC \\
20357      & A.~Zijlstra            & Sloan et al.\ (2012)                            & Carbon stars in other Local Group dwarf galaxies \\
\hline 
\end{tabular} 
\label{Tab-Programs}
\end{table*}

Most of the spectra considered here were obtained with the
low-resolution modules of the IRS:  Short-Low (SL), which
covers the 5.1--14.2~\mum\ range, and Long-Low (LL),
which covers 14.0--37.0~\mum.  Both modules have a
resolution ($\lambda$/$\Delta$$\lambda$) of $\sim$60--100.
For some of the fainter sources, spectra were obtained using 
only SL.  Program 40650 (Gruendl et al.\ 2008) observed 
several deeply embedded carbon-rich sources with SL and the
two modules with higher spectral resolution, Short-High (SH) 
and Long-High (LH).  They referred to these sources as
extremely red objects (EROs), and we have done the same.

All observations utilized the standard IRS nodding mode,
which produces spectra of the source in two positions in
each slit.  Each module has two apertures, which produce
spectra in two orders, along with a short `bonus' order,
which produces a short piece of the first-order spectrum
when the source is in the second-order aperture.  Thus,
to generate a full low-resolution IRS spectrum, eight
separate pointings of the telescope are required, and
these produce 12 spectral segments which must be combined.

The detailed nature of the observations varied substantially
among the samples considered here.  Some observers were
careful to match the integration times and number of
integration cycles in each aperture of a given module; others
prioritized reduced observing time over flexibility in
background subtraction.  Whenever possible, we differenced
images by aperture in SL, using the image with the source in
one aperture as the background for the image when the source
is in the other aperture.  For LL, we generally used the
image with the source in one nod as the background for an
image with the source in the other nod in the same aperture.
However, when forced to deviate from this default due to
background gradients, other sources in the image, or the
design of the observation, we shifted to whatever
background had the least structure.
Difference images still contain pixels which cannot be
calibrated (rogue pixels), which we replaced using the
{\sc imclean} algorithm (see Sloan \& Ludovici (2012) for 
more details).

Spectra were extracted from images using two methods:
tapered-column and optimal extraction.  The former sums the
spectra within a range of fractional pixels close to the
spectral trace.  The algorithms available with the SMART and
CUPID packages are functionally equivalent;\footnote{The
Spectroscopy Modelling Analysis and Reduction Tool (Higdon et 
al.\ 2004), and the Customizable User Pipeline for IRS
Data, available from the SSC.} we used CUPID.  Optimal
extraction fits a measured point-spread function (PSF) to the
image at each wavelength to minimize the noise.  We used the
algorithm available in SMART (see Lebouteiller et al.\ 2010 
for details) but ran it offline.

For tapered-column extraction, spectra were extracted from
individual images and then co-added.  For optimal extraction,
we coadded the images first to improve our ability to
properly locate faint sources, then extracted.

The tapered-column extractions are preferred in only a few
cases.  If the source is extended, optimal extraction 
produces artefacts due to the impossibility of fitting a PSF.  
For bright sources, the gain in signal-to-noise ratio (SNR) with 
the optimal extraction is negligible, and for the highest 
SNR cases, artefacts due to limits in our understanding of the 
PSF can be seen.  We used optimal extraction in most cases; 
it can improve the SNR by a factor of nearly two.  Only when
the spectra produced different structure did we rely on the 
tapered-column extraction.\footnote{Tapered-column extraction 
was kept for the following sources:  MSX LMC 787, IRAS 04374, 
IRAS 05568, MH~6 and W61-6-24}  Most of the spectra 
considered here are  publicly available from the CASSIS 
website, which provides both tapered-column and optimal 
extractions and guidance as to which is preferred in 
individual cases.

Spectra observed with SH and LH were extracted using
full-slit extraction, which is also available in SMART. 
Only the EROs from Program 40650 (Gruendl et al.\ 2008) are 
affected.

Another difference between observations was the requested
accuracy of the peak-up algorithm used to centre the
source in the spectroscopic slit.  Some of the spectra did
not utilize the highest accuracy, and as a result these
were more likely to suffer from partial truncation of the
source by the slit edges.  Given the narrow size of the
SL slit (3.6\arcsec) compared to the typical pointing
accuracy of $\sim$0.4\arcsec, any of the SL data could
be affected.  Sloan \& Ludovici (2012) found that for most 
observations, multiplying a spectral segment by a scalar 
would solve this pointing-induced throughput problem to 
within $\sim$2\%.

Throughput problems generally result in small discontinuities
between spectral segments.  We assumed that all corrections
should be up, to the best-centred spectral segment. 

We calibrated the spectra using HR~6348 (K0~III) as a 
standard star, supplemented with HD~173511 (K5~III) for 
LL data taken after the change in detector settings for that
instrument (starting with IRS Campaign 45).  Sloan et al.\
(2015) describe how the truth spectra for these sources were
constructed, tested, and cross-calibrated with other systems. 

\subsection{Considering the sample} 

Not all programs considered here observed AGB stars and RSGs 
exclusively.  As with G09, targets were selected from these 
programs by examining the IRS spectra, collecting additional 
photometry (see below), consulting SIMBAD and the papers 
describing these programs, and considering the results of the
radiative transfer modelling (see Sect.~\ref{themodel}).  

Our classification leads to a sample with 225 C stars and 
171 M stars, with the M stars including ten objects in the
foreground of the LMC and one in front of the SMC.  Our
C stars include R CrB stars and post-AGB objects which
other groups place in separate categories.  For example,
we count 160 stars in the LMC and 46 in the SMC, while
Sloan et al.\ (2016) count 144 and 40, respectively.  The
classifications from the SAGE Team
include 145 C stars in the LMC and 39 in the SMC (Jones et 
al.\ 2017b; Ruffle et al.\ 2015).  We have classified one
unusual object, 2MASS J00445256$-$7318258 (j004452) as a 
C star, due to the deep C$_2$H$_2$ absorption band centred 
at 13.7~\mum.  Ruffle et al.\ (2015) classify it as O-rich
due to the presence of strong crystalline emission features
at 23, 28, and 33~\mum.  Kraemer et al.\ (2017; They refer to the object as MSX SMC 049.) confirmed
its optical C-rich spectrum, noted its dual C/O chemistry
in the mid-infrared, and suggested that it may be a post-AGB object.

\section{Ancillary data} 

Tables~\ref{Tab-Csample} and \ref{Tab-Msample} in the 
Appendix list basic information for the C  and M stars 
respectively, including name, position, an abbreviated 
identifier used in subsequent tables and figures, the 
pulsation period, pulsation (semi-)amplitude and filter 
(see details in Section~\ref{pulsationper} below), and
remarks.  These tables and some of the figures use
the following classifiers for the oxygen-rich stars:
FG=Foreground; SG=Supergiant; MA=M-type AGB-star.

\subsection{Photometry} 

For all stars additional broadband photometry ranging 
from the optical to the mid-IR was collected from the 
literature, primarily using
VizieR\footnote{http://vizier.u-strasbg.fr/viz-bin/VizieR} 
and the NASA/IPAC Infrared Science
Archive\footnote{http://irsa.ipac.caltech.edu/}, using the 
coordinates given in Tables~\ref{Tab-Csample} and ~\ref{Tab-Msample}.  

In the optical we collected $UBVI$ data from Zaritsky et al.\ 
(2002, 2004) for the LMC and SMC,  
$UBVR$ data from Massey (2002) for the MCs, $BVRI$ data from 
Oestreicher et al.\ (1997) for RSGs in the LMC, 
OGLE-{\sc iii} $VI$ mean magnitudes from Udalski et al.\ (2008a,b),
EROS $BR$ mean magnitudes from Kim et al.\ (2014) and Spano et al.\ (2011),
MACHO $BR$ mean magnitudes from Fraser et al. (2008), 
and $VRI$ data from Wood, Bessell \& Fox\ (1983, hereafter WBF).

Near-infrared photometry comes from DENIS $IJK$ data 
from Cioni et al.\ (2000) and their third data release (The 
DENIS consortium 2005), the all-sky $JHK$ release of 2MASS 
(Skrutskie et al.\ 2006), the extended mission long-exposure 
release (2MASS-6X, Cutri et al.\ 2006), $JHK$ data from the 
IRSF survey (Kato et al.\ 2007), $JHK$ data from the LMC 
near-infrared synoptic survey (Macri et al.\ 2015),
SAAO $JHKL$ data from Whitelock et al.\ (1989, 2003),
and CASPIR $JHKL$ data specifically taken for the IRS 
observations (Sloan et al.\ 2006, 2008, Groenewegen et al.\ 
2007), and from Wood et al.\ (1992), and Wood (1998).
VMC data (Cioni et al.\ 2011) were used for selected 
very red sources, mostly to try to determine their pulsation 
periods (Sect.~\ref{pulsationper}).

Mid-infrared data include {\it IRAS} data from the Point 
Source Catalogue and the Faint Source Catalogue (Moshir et al.\ 
1989; Loup et al.\ 1997; only data of the highest quality 
flag were considered), photometry at 3.6, 4.5, 5.8, 8.0 
and 24~\mum\ from the SAGE survey of the LMC (Meixner et al.\ 
2006; two epochs), the S$^3$MC survey of the SMC, (Bolatto et 
al.\ 2007), and the SAGE-SMC catalogue (Gordon et al.\ 2011; 
also two epochs).  For selected very red sources we also used 
{\it Spitzer} photometric data from Gruendl et al.\ (2008) and 
Whitney et al. (2008).
We also used data from the {\it Wide-field Infrared Survey 
Explorer} ({\it WISE}; Wright et al.\ 2010), specifically 
from the {\it AllWISE} catalogue (Cutri et al.\ 2013), as well 
as data from the {\it Akari} Point-Source Catalogueue (up to five 
filters between 3 and 24~\mum) and Infrared Camera catalogue 
(IRC; 9 and 11~\mum) (Ishihara et al.\ 2010, Ita et al.\ 
2010, Kato et al.\ 2012).

Far-infrared data were obtained from the Akari Far-Infrared 
Surveyor (FIS, with four filters between 65 and 160~\mum; 
Yamamura et al.\ 2010), MIPS 70~\mum\ data from the SAGE survey,
and the Heritage survey, which included photometry from
{\it Herschel} at 70, 100, 250, 350, and 500~\mum\ (Meixner 
et al.\ 2013).  Reliable data beyond 60~\mum\ were available 
for only about a dozen C stars and  a half dozen O stars, 
mostly SAGE data at 70~\mum.  Additionally, MIPS-SED spectra 
(van Loon et al.\ 2010a,b) were used for four 
sources\footnote{With our designations: wohg64, bmbb75, 
msxlmc349, and iras05329.}.

The literature considered is not exhaustive, but it does 
include all recent survey data available in the near- and 
mid-IR, where these stars emit most of their energy.  
When fitting radiative transfer models, all data were 
considered as individual measurements with their reported 
errors.  No attempt was made to combine or average multiple 
observations in a given photometric band, or in similar bands 
from different telescopes.  Variability is an important 
characteristic of AGB stars and can influence the constructed 
SED and the fitting.  Fortunately, in the optical where the 
amplitude of variability is largest, mean magnitudes are 
available from the OGLE, MACHO and EROS surveys, with small 
errors on the mean magnitude.

In most cases, spectra from the IRS and the infrared 
photometry agreed.  In the 48 cases where they did not
($\sim$ 12\% of the sample), the spectra were scaled 
to the available photometry in wavelengths covered by the spectrum.  
In half of those cases, the correction was less than 25\%, 
and in 14 cases it was larger than a factor of two.

\subsection{Pulsation periods} 
\label{pulsationper}

An extensive effort was made to obtain pulsation periods for 
the sample. The most important data sources are the OGLE, 
EROS and MACHO photometric surveys. Although periods have 
been published for LPVs from these surveys, we downloaded the 
original data and derived periods independently using the 
publicly available code {\it Period04} (Lenz \& Breger 2005).
OGLE-III $I$-band data are available through the OGLE Catalogue 
of Variable 
Stars.\footnote{ftp://ftp.astrouw.edu.pl/ogle/ogle3/OIII-CVS/}
The EROS-2 data for more than 150~000 stars that were used by 
Kim et al.\ (2014) are available 
online,\footnote{http://stardb.yonsei.ac.kr/} and the data 
for the few stars that were missing were kindly provided by 
Dr.\ Jean-Baptiste Marquette (private communication).  The 
correspondence between EROS-2 identifier and coordinates is 
provided by a separate database, but is also available 
through SIMBAD and VizieR.  MACHO data are also 
online.\footnote{http://macho.nci.org.au/}
In this case any association listed in SIMBAD and VizieR 
between our sources and the MACHO counterpart was only taken 
as guidance, and we independently searched for all MACHO 
targets within 2.5\arcsec\ of our targets. 

Other sources of optical data were also considered, 
specifically the Catalina Sky Survey (CSS, Drake et al.\ 
2014),\footnote{http://nunuku.caltech.edu/cgi-bin/getcssconedb\_release\_img.cgi}
ASAS-3 data (Pojmanski 
2002)\footnote{http://www.astrouw.edu.pl/asas/?page=aasc\&catsrc=asas3},
and data from the Optical Monitor Camera (OMC) onboard {\it 
INTEGRAL} (Mas-Hesse et al.\ 
2011).\footnote{https://sdc.cab.inta-csic.es/omc/secure/form\_busqueda.jsp}

Many of the redder sources have traditionally been monitored 
in the near-infrared (for example Wood 1998, Whitelock et al.\ 
2003), and these data have also been used, either taking 
directly the quoted periods and amplitudes or in some cases
combining the data and rederiving the periods.  For red stars 
with no previous period determination or where the available 
infrared data were sparse, the VMC database (Cioni et al.\ 
2011) was consulted.  In the $K$-band the VMC observations 
typically have 10--15 data points spread over a relatively 
short timespan (6--12 months), but when combined with other 
data, even if from a single epoch, a reliable and unique 
period could be derived in many cases (Groenewegen et al.\ in prep.)

To analyse the variability of the more embedded sources, 
we have used the data from {\it WISE} differently than
described in the previous section, where the focus was on 
assembling photometric data to be fitted with radiative
transfer models.  To investigate variability, we have 
followed Sloan et al.\ (2016) and used the {\it AllWISE 
Multiepoch Photometry Table} and the {\it NEOWISE-R Single 
Exposure (L1b) Source Table} (from the NEOWISE reactivation 
mission; Mainzer et al.\ 2014) but with some changes to
their method.  We did not average together data taken 
within a few days of each other.  We considered data up 
to and including the June 2017 release, which gives two more
years of data than were considered before.  We also focussed 
just on the W2 filter at 4.6~\mum, as it is brighter than W1 
for the reddest sources.

When data were available, we combined the W2 data with IRAC 
4.5~\mum\ data from the {\it SAGE-VAR} catalogue (Riebel et 
al.\ 2015), which adds four epochs obtained during the 
warm {\it Spitzer} mission on the Bar of the LMC and the core 
of the SMC with the original epochs obtained during the {\it SAGE} 
and {\it SAGE-SMC} surveys.  The effective wavelengths 
of the W2 and IRAC 4.5 filters are similar, but for the 
reddest sources for which these data were used, the 
difference can be of the order of a few tenths of a 
magnitude.  To shift the W2 at 4.6~\mum\ to 4.5~\mum, 
we did not use the colour corrections derived by Sloan et 
al.\ (2016).  Instead, we used the radiative transfer models 
fitted to the photometry to determine the proper adjustment 
from the WISE to the IRAC filter.

Tables~\ref{Tab-Csample} and \ref{Tab-Msample} in the 
Appendix list the adopted pulsation period, the amplitude 
(in the mathematical sense, sometimes referred to as the 
semi-amplitude, or in other words, half the peak-to-peak 
amplitude), the filter, and the reference to the data used.  

The sample includes about 180 stars with periods listed by OGLE.  
In 85\% (88\%) of the cases the period we derive from
the OGLE data agrees within 5\% (10\%) with the first of
three possible periods listed by OGLE.  In an additional 6\%
of the cases our adopted pulsation period corresponds to the
second period listed by OGLE.  There are also a few cases
where the period we find is about double that of the first
OGLE period.  The final periods listed in the tables also
include the analysis from EROS and MACHO data when available.

As mentioned above, Sloan et al. (2016) used  {\it AllWISE}
and {\it NEOWISE} data to derive previously unknown periods
for five C stars (their Fig.~20).  For three stars we quote
periods here based on $K$-band data from the VMC survey and
the literature (Groenewegen et al.\ in prep), that agree to
within 3\% with the periods found by Sloan et al.  For 
the other two stars, we independently derived the periods 
based on the W2 filter (combining it with [4.5] data for one 
object).  Despite the differences in our approaches, the 
periods agree within 1\%.

\section{The model} 
\label{themodel}

The models are based on the `More of DUSTY' (MoD) code 
(Groenewegen 2012) which uses a slightly updated and modified 
version of the {\em DUSTY} dust radiative transfer (RT) code 
(Ivezi\'c et al.\ 1999) as a subroutine within a minimization algorithm.  

\subsection{Running the radiative transfer code} 

The RT code determines the best-fitting 
dust optical depth, luminosity, temperature at the inner 
radius, $T_{\rm c}$, and index of the density distribution, 
$\rho \sim r^{-p}$ by fitting photometric data and spectra 
for a given model atmosphere and dust composition.  (The code 
can also consider visibility data and 1D intensity profiles, 
but these data are not available for the sample considered here.)  
Each of the four free parameters may be fixed or fitted in 
the RT code, see Section~\ref{bestmodel}.

The outer radius in the models is set to a value where the 
dust temperature reaches about 20~K. This implies values of 
$(4-22) \cdot 10^3$ times the inner radius, which correspond 
to outer radii of less than $\sim 25\arcsec$ in all cases.
However, depending on wavelength, the emission comes from a 
much smaller region.  As a test, the SED was calculated for 
one of the reddest sources with a large default outer radius 
of $13~000~R_{\rm in}$, and then re-run with progressively
smaller outer radii.  At 70~\mum\ the flux is reduced by 5\% 
when decreasing the outer radius to $2~000~R_{\rm in}$, 
corresponding to about 2\arcsec.  In comparison the FWHM of 
the MIPS 70~\mum\ band is 18\arcsec.  Emission at shorter 
wavelengths comes from an even more compact region, for example 
\less~1\arcsec\ at 24~\mum.  Generally, the PSFs of the 
combinations of instrument and filter of the datasets listed 
in Sect.~3 match the physical size of the emitting region at 
the distance of the MCs quite well.  The exception is the 
WISE 4 filter at 22~$\mu$m with a PSF of 12\arcsec, which 
is much larger then the size of the emitting region and makes
background subtraction more important.  In fact, this is the
data point that has been excluded most frequently from the 
SEDs, in 25 of the 370 sources for which it was available.

We masked those portions of the IRS spectra with poor S/N or 
those affected by background subtraction problems and did not 
include them in the minimization procedure.  In 
addition, spectral regions dominated by strong molecular 
features that are not included in the simple model 
atmospheres are also excluded for the C stars, that is the 
regions 5.0--6.2~\mum\ (CO + C$_3$, e.g.\ J\o rgensen et 
al.\ 2000), 6.6--8.5~\mum\ and 13.5--13.9~\mum\ (C$_2$H$_2$, 
e.g.\ Matsuura et al.\ 2006).  No regions were excluded in 
the fitting of the O-rich stars.

The photospheric models for C stars come from Aringer et 
al.\ (2009)\footnote{
\url{http://stev.oapd.inaf.it/synphot/Cstars/}}, while the 
M stars are modelled by a MARCS stellar photosphere model 
(Gustafsson et al.\ 2008)\footnote{\url{ 
http://marcs.astro.uu.se/}}.  For the C stars, the models 
assume 1/3 solar metallicity for the LMC and 1/10 for the 
SMC and other Local Group galaxies.  For M stars, the 
corresponding values are $-0.5$ and $-0.75$ dex.  The 
model atmospheres also depend on $\log g$, mass, and C/O 
ratio (for the C-stars).  The SED fitting is not sensitive 
to these values, and we adopted fixed values $\log g= 0.0$ 
or $+0.5$, 1 or 2~\msol, and C/O= 1.4.

To determine luminosities from the RT output, distances 
of 50 kpc to the LMC and 61 kpc to the SMC are adopted, while 
for the other Local Group galaxies the distances adopted by 
Sloan et al.\ (2012) are kept:  84.7 kpc for Sculptor, 
104.7 kpc for Carina, 140.6 kpc for Fornax, and 259.4 kpc for Leo I.

The model results have been corrected for a typical 
$A_{\rm V}$ = 0.15 mag for all Magellanic stars, 
0.06 mag for Sculptor, 0.08 mag for Carina and 
Fornax, and 0.09 mag for Leo I.  We adopted the 
reddening law of Cardelli et al.\ (1989) for other 
wavelengths.  The exact value has little impact on the 
results, as it corresponds to \less 0.02 mag of extinction 
in the near-IR.

\subsection{Dust grain properties} 

The dust around the C stars is assumed to be a combination of 
amorphous carbon (AMC), silicon carbide (SiC), and magnesium 
sulphide (MgS).  The optical constants are taken from Zubko et 
al.\ (1996, the ACAR species, denoted `zubko' in 
Table~\ref{Tab-Cstar}) for AMC, Pitman et al.\ (2008, denoted 
`Pitm' in Table~\ref{Tab-Cstar}) for SiC, and Hofmeister et 
al.\ (2003, denoted `Hofm' in Table~\ref{Tab-Cstar}) for MgS.

Models were run mostly for a single grain size 0.15~\mum, 
although some models were also explored with $a$= 0.10 and 
0.30~\mum. The current modelling does not allow us to 
determine the grain size (or even grain size distribution). 
Nanni et al.\ (2016) recently found that the near- and 
mid-IR colours of SMC C-stars can be described better with 
grains of size $\less$0.12~\mum\ than with grains $\more$0.2~\mum.

The models by G09 did not consider MgS, and the 22--39 $\mu$m 
wavelength range was specifically excluded in their fitting.  
Also, the absorption and scattering coefficients were 
calculated in the small-particle limit for spherical 
grains.  Here, MgS is included in the fitting, but spherical 
grains are known not to match the observed profile of MgS 
(Hony et al.\ 2002).
The absoption and scattering coefficients for MgS have been 
calculated using a distribution of form factors (Mutschke et 
al.\ 2009) close to the classical CDE (continous distribution 
of ellipsoids; see Min et al.\ 2006).  This distribution fits the 
observed shape of the 30~$\mu$m feature reasonably well.

The identification of MgS as the carrier of the 30~\mum\
feature by Goebel \& Moseley (1985) has come under some
scrutiny in the past few years.  The primary issue is that
the strength of the 30~\mum\ feature in some post-AGB objects
would violate abundance limits if the grains were solid MgS
(Zhang et al.\ 2009).  However, if MgS forms a layer on 
a pre-existing core, it will produce the observed feature 
without requiring too much Mg or S.  Sloan et al.\ (2014) 
reviewed the spectroscopic evidence supporting the case for 
layered grains, addressed other concerns about MgS as the 
carrier of the 30~\mum\ feature, and concluded that it 
remains the best candidate.

The  absorption and scattering coefficients for SiC and 
AMC are calculated using a distribution of hollow spheres 
(DHS, Min et al.\ 2003).  In DHS, the optical properties are 
averaged over a uniform distribution in volume fraction 
between zero and $f_{\rm max}$ of a vacuum core, where the 
material volume is kept constant, in order to simulate the 
fluffiness of real grains.  G09 found that $f_{\rm max} = 
0.7$ fitted the data well, and we use that value here.

AMC, SiC and MgS are then mixed in proportions in order to 
fit the data, that is the strength of the SiC and MgS feature.
This method assumes that the mixture is uniform throughout
the envelope and that the grains are in thermal contact.

For M stars the dust chemistry is richer than that for C stars.  
We considered several species, but found that we could model 
the observed data with four dust components: 
amorphous silicates (olivine or MgFeSiO$_4$; Dorschner et al.\ 1995; 
denoted `Oliv' in Table~\ref{Tab-Mstar}), amorphous alumina (Begemann et al.\ 1997; denoted `AlOx'),  
metallic iron (Pollack et al.\ 1994; denoted `Fe'), and 
crystalline forsterite (Mg$_{1.9}$Fe$_{0.1}$SiO$_4$; Fabian et al.\ 2001, denoted `Forst').

The absorption and scattering coefficients are calculated 
using DHS with $f_{\rm max} = 0.7$, and for grain sizes 0.1, 
0.2 and 0.5~\mum, compared to a more traditional value of
$\sim$0.1~\mum.  For some Galactic targets, large grains 
have proven necessary when near-IR visibility data are
available, because they provide an additional scattering 
component.  Examples include IRC~+10~216 and OH~26.5, which 
were fitted with MoD code by Groenewegen et al.\ (2012) and 
Groenewegen (2012), respectively. 
Recently, Norris et al. (2012), Scicluna et al.\ (2015), and 
Ohnaka et al.\ (2016) found evidence for grains in the range 
0.3--0.5~\mum, and 0.1--1~\mum\ grains are advocated to drive 
the outlow around oxygen-rich stars (H\"ofner 2008, Bladh \& 
H\"ofner 2012).

\subsection{Finding the best model} 
\label{bestmodel}
 
The MoD code determines the best-fitting luminosity, dust 
optical depth, temperature at the inner radius, and index of 
the density distribution.  Groenewegen (2012) describes how
to identify the best-fitting model by minimizing a $\chi^2$ 
parameter.

The luminosity is fitted in all cases.  Whether the other 
three parameters have been fitted or fixed is indicated in 
Tables~\ref{Tab-Cstar} and \ref{Tab-Mstar}.  In about 8\% of 
the objects, the data show no evidence of a dust excess, and 
the dust optical depth is fixed to a small value.
In about one-third of the C-stars and half of the O-stars, 
the dust optical depth is the only additional free parameter.
When not fitted, $T_{\rm c}$ is fixed to the typical 
condensation temperature (1200~K for C-stars and 1000~K for 
O-stars in most cases) and $p$ is fixed to two, that is the density law for a 
constant outflow velocity and MLR.

Increasing the number of free parameters will always decrease 
the reduced $\chi^2$ of the model.  To avoid overfitting and 
penalizing the addition of free parameters, MoD also 
calculates the Bayesian information criterion (BIC; Schwarz 
1978; see also Groenewegen 2012).  The best-fitting model is 
the one with the lowest BIC value.

Some of the parameters that are determined are external to 
MoD, as they are available on a discrete grid only.  For
example, the model atmospheres (both MARCS and Aringer et 
al.\ 2009) are available on a grid with 100~K spacing.  In
addition, the grain absorption and scattering properties are
pre-calculated for a discrete number of grain sizes and dust
compositions.  For C stars, we considered AMC, SiC and MgS 
in ratios 100 : $x$ : $y$, where $x$ and $y$ are integer 
multiples of five.  We treated O-rich stars with a similar grid
for silicates, alumina, metallic iron and forsterite. 

MoD is then run over the grid of effective temperatures (for 
a typical dust composition) to find the best-fitting model
atmosphere.  Then the model atmosphere is fixed to that 
effective temperature, and the model is run over the grid of 
grain sizes and dust compositions to find the best dust 
model.  The input and output of the models (parameters, 
reduced $\chi^2$, BIC) are stored, so that the lowest BIC 
(the best-fitting model) may be found, and errors estimated 
for the parameters.

\section{Results} 

Tables~\ref{Tab-Cstar} and \ref{Tab-Mstar} in the Appendix 
list the parameters of the models which best fit the observed 
data for the C stars and M stars, respectively.  
Figures~\ref{Fig-Cstars} and \ref{Fig-Mstars} in the Appendix 
show the best-fitting model to the SED and IRS spectra for 
these two groups.  The tables include the luminosity and 
MLR, calculated assuming an expansion velocity v$_{\rm exp}$ 
of 10~\ks\ and a gas-to-dust ratio of 200.  Observations of
CO line widths suggest that the outflow velocity may depend 
on luminosity and metallicity (e.g.\ van Loon et al.\ 2000, 
Lagadec et al.\ 2010, Groenewegen et al.\ 2016; Goldman et 
al.\ 2017).  The gas-to-dust ratio in the circumstellar 
outflows is expected to scale with the initial metallicity 
for O rich sources, but this is not clear for C rich sources,
which fuse carbon in their interiors.  In both cases, though, 
the nature of that dependence is uncertain, and we hold both 
velocity and gas-to-dust ratio constant to aid comparison.  
The MLRs are proportional to both quantities and can be 
scaled as desired.

The fitting routine also provides uncertainties for the MLR, 
dust temperature at the inner radius, and luminosity.  These 
are typically small, of order 1\%, and are not reported. 
The realistic 1$\sigma$ errors are larger, and can be 
estimated from a comparison of model runs with different 
parameters and different realizations of the 
photometric data.  They are typically 10\% in luminosity, 
25\% in MLR and 50~K in $T_{\rm c}$,   
and have been estimated from the spread in the parameters in 
all models that have a $\chi^2$ less than twice the value of 
the best-fitting model.
The difference between the small fitting error 
and the realistic error is likely due to the difficult 
treatment of source variability in the fitting.  Currently, 
multiple measurements at a given wavelength have been 
assigned their respective photometric errors.  However, 
the source variability, especially in Miras and long-period 
variables is usually much larger than the photometric 
uncertainty.  The fitting will lead to a larger best-fit 
$\chi^2$ than if the same source were only fitted with, say, 
a single $K$-band photometric point.

\subsection{Colour relations} 

\begin{figure}
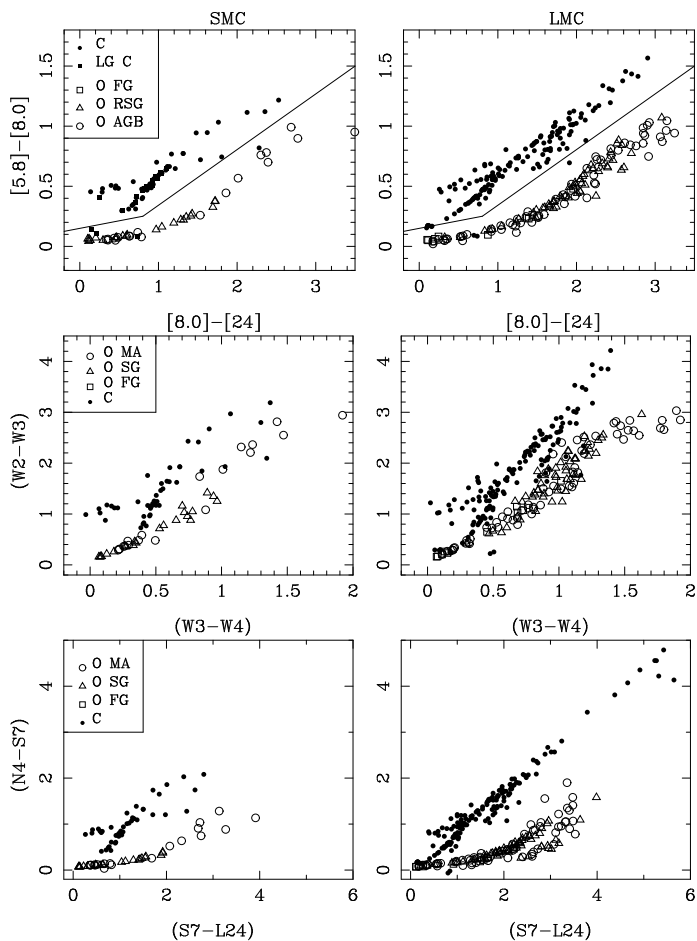


\begin{minipage}{0.49\textwidth}
\resizebox{\hsize}{!}{\includegraphics{sirtf_ccd4_2017.ps}}
\end{minipage}

\begin{minipage}{0.49\textwidth}
\resizebox{\hsize}{!}{\includegraphics{sirtf_ccd5_2017.ps}}
\end{minipage}

\begin{minipage}{0.49\textwidth}
\resizebox{\hsize}{!}{\includegraphics{sirtf_ccd7_2017.ps}}
\end{minipage}

\caption[]{Colour-colour diagrams for the SMC (left-hand panels) 
and LMC (right-hand panels).  Top panels:  the combination of 
IRAC and MIPS colours from {\it Spitzer}. 
Middle panels: {\it WISE} colours.  
Bottom panels:  {\it Akari} colours.  
Symbols are explained in the legend.
The lines indicate the border between C- and O-rich stars 
(see text).}
\label{Fig-CCD} 
\end{figure}

Some of the analysis presented by G09 can be improved here 
thanks to the larger sample size.  Figure~\ref{Fig-CCD} shows 
a synthetic colour-colour diagram (CCD) generated from the
best-fitting models.  Plotting [5.8]$-$[8.0] vs.\ 
[8.0]$-$[24] is effective at distinguishing M from C 
stars (Kastner et al.\ 2008, G09).  The larger sample only 
reinforces this conclusion.  Following this result, we also 
constructed CCDs with similar colours generated by using 
{\it WISE} and {\it Akari} data.  For {\it WISE} data, 
W2$-$W3 vs.\ W3$-$W4 has the most discriminatory power, and 
for {\it Akari}, it is N4$-$S7 vs.\ S7$-$L24.  Both CCDs show 
behaviour similar to the {\it Spitzer}-based CCD, but they do 
not separate the M and C stars quite as well.

\begin{figure}
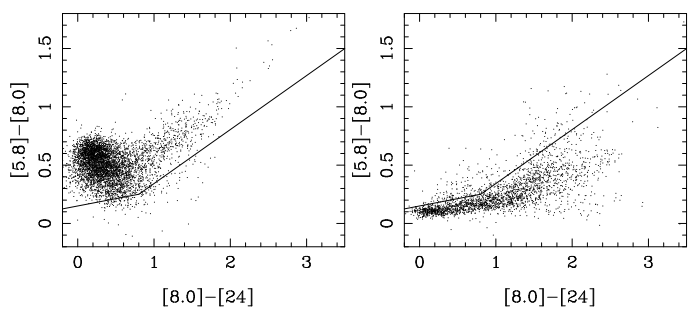


\begin{minipage}{0.24\textwidth}
\resizebox{\hsize}{!}{\includegraphics{CCDRiebel_C_paper.ps}}
\end{minipage}
\begin{minipage}{0.24\textwidth}
\resizebox{\hsize}{!}{\includegraphics{CCDRiebel_O_paper.ps}}
\end{minipage}

\caption[]{Colour-colour diagrams for the LMC for C-rich 
(left-hand panel) and O-rich stars (right-hand panel), as 
classified by Riebel et al.\ (2012).  The lines indicates the 
border between C- and O-rich from Fig.~\ref{Fig-CCD}.}
\label{Fig-CCDRiebel} 
\end{figure}

Each of the top panels in Fig.~\ref{Fig-CCD} also 
includes two lines that separate M and C stars, with a break 
at [5.8]$-$[8.0] = 0.8 mag).  To the blue, the M stars 
are those with [5.8]$-$[8.0] $< 0.125\, ([8]$$-$$[24]) + 
0.150$.  To the red, they follow [5.8]$-$[8.0] $< 0.463\, 
([8]$$-$$[24]) - 0.121$.  Over 95\% of the stars classified as 
C stars appear above this line.  Ten stars classified as 
C stars, however, stray into the M-star territory.  

Two of the strays are red, with [8.0]$-$[24.0] at 1.9 mag (in the LMC) 
and 2.3 mag (in the SMC).  The red 
SMC object, IRAS~00350, is classified by Sloan et al.\ (2014) 
as a C-rich post-AGB object and is evolving to the red at 
shorter wavelengths, as expected for more evolved post-AGB 
objects and young planetary nebulae.  The red LMC source is 
MSX~LMC~663, which shows a peculiar SED and the highest 
effective temperature of the C-rich sample.  It has not
actually crossed into the colour range defined by the sequence
of M stars, and the fact that it has crossed the boundary 
line illustrates that one could easily draw the boundary 
slightly differently.  A few other C-rich objects have also 
strayed toward the sequence of M stars.

The remaining eight C stars below the boundary with M stars
are all relatively blue and dust-free.  Three are in the
Carina dSph, and their spectra showed no evidence of 
amorphous carbon or SiC dust (Sloan et al.\ 2012).  Two,
HV~5680 and WBP~17, have clearly C-rich optical spectra and
have infrared spectra showing little or no amorphous carbon
dust (Sloan et al.\ 2008).  The remaining three are from the
SAGE sample observed by the IRS.  Woods et al.\ (2011) 
classified two, PMP~337 and KDM~6486, as 
`C-AGB', and they are classified by Sloan et al.\ 
(2016) as naked and nearly naked C stars (respectively).  
Woods et al.\ (2011) classified the third source, 
LMC-BM~11-19, simply as `star', and it did not appear 
in the C-rich sample of Sloan et al.\ (2016).  All ten of 
these stars are close to naked stars, and all dust-free 
stars, whether they be C-rich or O-rich, will fall in the 
same region in most infrared colour-colour spaces, making some 
overlap inevitable for the bluest sources.

To test the robustness of how the C stars and M stars 
separate by colour, we examined the sample of stars defined by 
Riebel et al.\ (2012).  Figure~\ref{Fig-CCDRiebel} plots 
those stars with relative errors at 24~\mum\ $<$ 5\% and 
classified by them as C-rich or O-rich.  Of the 4867 C-rich 
stars, only 161, or 3.3\%, stray into the zone of the M stars 
as defined above.  Of the 2496 stars classified as O-rich, 
2233, or 89\%, fall in the O-rich zone.  These high 
percentages testify to the discriminatory power of this CCD.

\subsection{Bolometric corrections} 
\label{bc}

\begin{figure*}
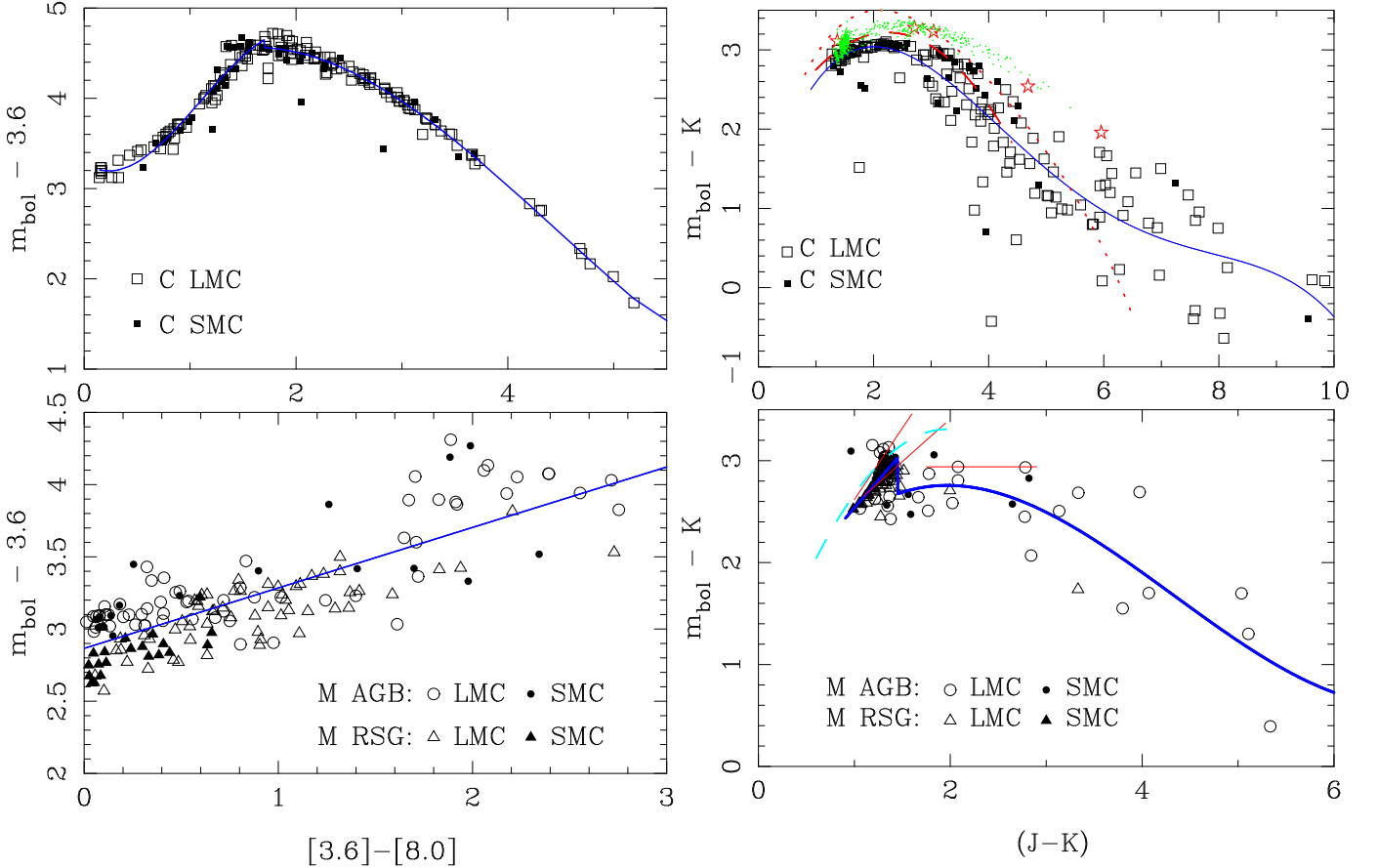


\begin{minipage}{0.49\textwidth}
\resizebox{\hsize}{!}{\includegraphics{BC_col_2017.ps}}
\end{minipage}
\begin{minipage}{0.49\textwidth}
\resizebox{\hsize}{!}{\includegraphics{BC_col1_2017.ps}}
\end{minipage}

\caption[]{Bolometric correction at 3.6~\mum\ versus
[3.6]$-$[8.0] colour (left) and $K$ versus $J-K$ colour (right)
for C stars (top panels) and M stars (bottom panels).  
The solid dark-blue lines are polynomial fits to the data
(see Table~\ref{Tab-Coef}).  See Section~\ref{bc} for an 
explanation of sources not fitted or plotted.  
For C stars in the top right panel, red stars indicate models 
by Nowotny et al.\ (2013), green dots indicate models by 
Eriksson (2014),  the dotted red line indicates the fit 
by Kerschbaum et al.\ (2010),
and the dashed red line indicates the fit by Whitelock 
et al.\ (2006).
For M stars in the lower-right panel, the straight red 
lines indicate relations "A", "B", and "C" from Kerschbaum 
et al.\ (2010), valid in the range $1.0 < J-K< 1.6$, $1.1 < 
J-K < 1.95$, and $1.75 < J-K< 2.9$~mag, respectively.
The dashed light blue line indicates the fit by WBF.
}
\label{Fig-BC}
\end{figure*}

Figure~\ref{Fig-BC} shows the bolometric correction (BC) at 
3.6~\mum\ versus [3.6]$-$[8.0] colour, and at $K$ versus 
$(J-K)$ colour (in the 2MASS system) for the synthetic colours
determined from the best-fitting models for all sources.  The 
bolometric magnitude $M_{\rm bol}$ has been calculated from 
$-2.5 \, \log L_{\rm bol}/L_{\odot} +4.72$~mag.  The data 
have been fitted by polynomials, and Table~\ref{Tab-Coef} 
lists the coefficients and the rms in the fit.  Relations 
like these make it possible to estimate bolometric magnitudes 
with an estimated uncertainty of about 0.1--0.3~mag, which 
should be sufficient for most applications.  Such an estimate 
could also serve as a first guess in a more detailed automated modelling.

The BC for C stars at 3.6~\mum\ is the best defined relation 
and has an rms scatter below 0.1~mag.  Polynomials are 
fitted separately to the data on either side of [3.6]$-$[8.0] 
= 1.7~mag, and they exclude six stars outside the plotted colour 
range as well as three SMC stars with [3.6]$-$[8.0] colours in 
the range 1.2--2.8~mag that are below the line (from
left to right:  CV 78, IRAS 00350 and j004452).

For the C stars, the BC for $(J-K)$ agrees well with 
the quadratic relation by Kerschbaum et al.\ (2010) in the 
range in common ($J-K < 4$).  Our values are on the low side 
compared to observations by Whitelock et al.\ (2006) and 
models by Nowotny et al.\ (2013) and Eriksson et al.\ (2014).
The fitting excludes 24 C stars with $(J-K)$ $>$ 10~mag
as well as the outliers SAGEMCJ054546 (near $J-K \sim$ 1.75~mag) 
and IRAS 05278 (near $J-K \sim$ 4.05~mag).

For the redder carbon stars, the scatter about the fitted 
polynomials is substantially larger for the bolometric 
corrections based on $(J-K)$ compared to [3.6]$-$[8.0], 
primarily because [3.6]$-$[8.0] better samples the peak of 
the SED for these sources.  Another contribution to the 
scatter at $(J-K)$ is that more evolved carbon stars can show
excesses of blue radiation, which Sloan et al.\ (2016) 
attribute to scattered light escaping from shells as they
begin to depart from spherical symmetry (see 
Section~\ref{sec-mlr}).  For all of these reasons, we would 
recommend the use of BCs based on $(J-K)$ only for $J-K 
\lesssim$~2~mag.  Beyond that limit, BCs based on colours from 
longer-wavelength filters will be more reliable.

The BCs for M stars based on [3.6]$-$[8.0] show a
well-defined relation with a scatter of about 0.2~mag. 
No data were excluded from the plot or the fitted
polynomial.

We have fitted two polynomials to estimate the BC for M 
stars based on $(J-K)$, with a break at $J-K$ = 1.45~mag.  One M 
star (IRAS 05329 at $J-K \sim$ 7.1) is neither plotted nor 
fitted.  In the regime $J-K <$ 1.45~mag four stars are excluded 
by sigma-clipping (HV~12122, HV~12070, NGC~1948 WBT~2215, 
SAGEMCJ050759).  The break at $J-K$ = 1.45~mag results from a
clear dichotomy in the data, which can also be seen the data
presented by Kerschbaum et al.\ (2010).  For bluer colours, 
where most of the stars are located, a simple straight line 
fits well, with a dispersion less than 0.1~mag.  For 
redder colours a third-order polynomial is fitted, but the 
data show much more dispersion.  As for the carbon stars,
the BC based on [3.6]$-$[8.0] is much better behaved for the
redder sources in the sample.  Here the shift away from
BCs based on $(J-K)$ should occur at $J-K \sim$~1.45~mag.

\begin{table*}

  \caption{Bolometric corrections fitted to the data.}
  \begin{tabular}{lrrrrrrr}
  \hline
         Condition         &  $a_0$   & $a_1$       & $a_2$     & $a_3$    & $a_4$       &  rms \\
\hline
C stars 0.0 $<$ [3.6]-[8.0] $<$ 1.7 & 3.290 & $-0.843$ &  +1.99307  & $-0.604618$ &       & 0.063  \\
C stars 1.7 $<$ [3.6]-[8.0] $<$ 5.5 & 3.386 &  +1.565  & $-0.58910$ &  +0.043892  &       & 0.073  \\

M stars 0.0 $<$ [3.6]-[8.0] $<$ 3.0  & 2.866  & $+0.419$ &             &              &       & 0.21 \\ 

C stars 0.9 $<$ ($J-K$) $<$ 10    & 0.919  & +2.482 & $-0.91577$ & +0.111553 & $-0.004608$ &   0.36 \\ 

M stars  0.9 $<$ ($J-K$) $\le$ 1.45  & 1.453      & +1.084   &            &              &      &  0.096  \\ 
M stars 1.45 $<$ ($J-K$) $<$ 6.0 & 2.354      & +0.453   & $-0.13580$  &              &      &  0.31   \\ 

\hline
\end{tabular}
\tablefoot{Bolometric corrections to $K$ and [3.6] are
computed as BC= $\sum_{i}^{} \; a_i \; [{\rm colour}]^{i}$.
}

\label{Tab-Coef}
\end{table*}

\subsection{Mass-loss rates} 
\label{sec-mlr}

\begin{figure}  
\begin{minipage}{0.49\textwidth}
\resizebox{\hsize}{!}{\includegraphics{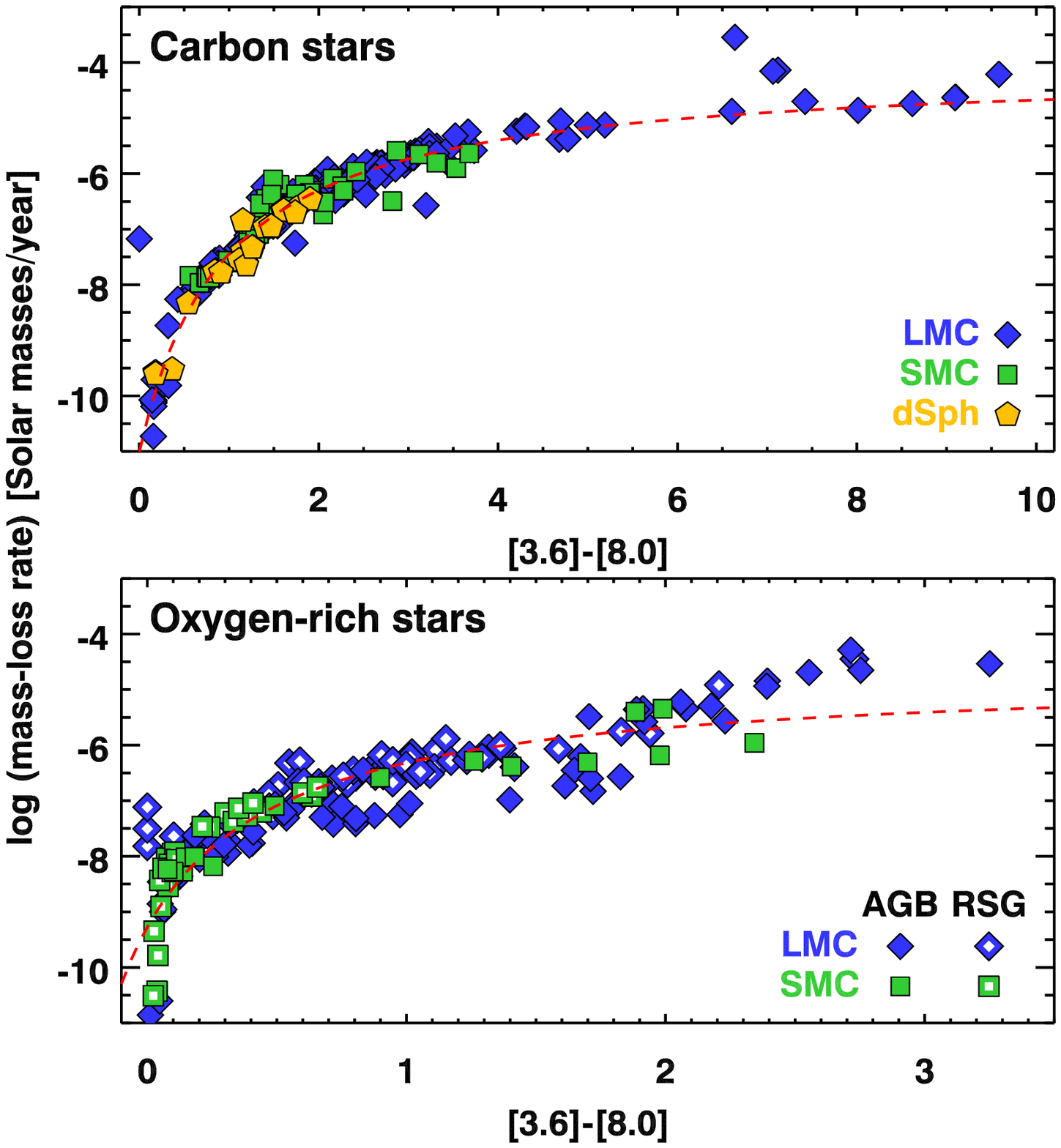}}
\end{minipage}

\caption[]{Mass-loss rate versus colour for C stars (top 
panel), and M stars (bottom panel).  The fitted relations
are shown as dashed orange curves and are given by equations
(1) and (2).  The carbon stars include sources from the LMC,
SMC, and dwarf spheroidal galaxies.}
\label{Fig-MLCol} 
\end{figure}

Figure~\ref{Fig-MLCol} plots MLR as a function of 
[3.6]$-$[8.0] colour generated from synthetic photometry 
of the fitted models.  It closely resembles the corresponding 
figure from G09.  Generally, redder colours are associated 
with larger MLRs, as expected.  The relations are tight, with 
no clear dependence on metallicity (assuming that the 
expansion velocity and gas-to-dust ratio do not depend on 
metallicity).

Sloan et al.\ (2008) fitted a line to the logarithm of the 
MLR as a function of colour for carbon stars.  They used a 
spectroscopically derived colour ([6.4]$-$[9.3]), which 
closely follows the photometric [5.8]$-$[8] colour (Sloan et 
al.\ 2016).  However, the older sample did not include 
targets observed later in the {\it Spitzer} mission, which 
added more sources at the red and blue ends of the colour 
range.  For the carbon stars, the additional sources do not 
follow a single linear relation.  Matsuura et al.\ (2009) 
used a three-parameter function based on the inverse of the 
colour, which adds the necessary curvature.  For the carbon 
stars, 
\begin{equation}
  {\rm \log} \, \dot{M} = -4.080 \, - \, 
   \frac{6.531}{([3.6]-[8.0]) \, + \, 0.941}.
\end{equation}
The typical uncertainty in log (MLR) is about 0.22 dex.  The 
three data points with [3.6]$-$[8.0] $\sim$ 6.5~mag and 
$\log \, \dot{M} > -4$ are excluded, because they have
probably evolved off of the C-rich sequence defined by the 
rest of the sample.  Two of these sources appear in the 
sample of Magellanic carbon-rich objects of Sloan et al.\ 
(2014), primarily because their IRS spectra were redder than 
any of the carbon stars considered by Sloan et al.\ (2016).  
The spectra showed no other obvious spectral features 
associated with post-AGB objects, such as forbidden lines, 
fullerenes, or polycyclic aromatic hydrocarbons (PAHs).  The 
third source is one of the EROs from the sample by Gruendl et 
al.\ (2008), ERO~0518117.  

As a group, the observed photometry of these sources suggests
that they have begun to evolve off of the AGB, by showing
both reduced variability and bluer colours at shorter
wavelengths.  C stars generally show a tight relation between 
most infrared colours, so that in most CCDs, they fall along 
an easily recognisable sequence.  However, some of the 
reddest sources depart from that sequence.  For example, 
several of the EROs have a bluer colour at [3.6]$-$[4.5] than 
expected from [5.8]$-$[8.0] (Fig.~13 from Sloan et al.\ 
2016).  These sources may be deviating from spherical 
symmetry, allowing some light from the central star to escape 
via scattering in the poles of the extended envelope.  One 
should keep in mind that our models assume spherical 
symmetry.  Sloan et al.\ (2016, Fig.~14) also found that the 
variability of C stars increases to a [5.8]$-$[8.0] colour of 
$\sim$1.5~mag, but past that colour it decreases.  One 
would expect decreased variability as a star sheds the last 
of its envelope and departs the AGB.

For the M stars,
\begin{equation}
  {\rm \log}  \, \dot{M} = -4.708 \, - \, 
   \frac{2.488}{([3.6]-[8.0]) \, + \, 0.545}.
\end{equation}
The typical uncertainty is about 0.49 dex in 
$\log \, \dot{M}$, which is larger than for the C stars, due 
mostly to the apparent split in the relation for AGB stars 
and red supergiants, with the supergiants usually showing 
higher MLRs.  This relation is not valid at the bluest 
colours, [3.6]$-$[8] $\lesssim$ 0.1~mag, because the 
actual MLRs drop to zero much more quickly than the fitted 
relation indicates. 

The major difference with G09 is that the current C-rich 
models use the optical constants for AMC from Zubko et al.\ 
(1996), while G09 used the constants from Rouleau \& Martin 
(1991).

\section{Discussion} 

\subsection{Comparison to other works} 
\label{sect-comp}

\subsubsection{Groenewegen et al.\ (2009)} 

First we compare the MLRs derived by G09 to the present 
paper, which includes more photometry for the SED, uses 
different model atmospheres, and uses improved code for 
radiative transfer.  In addition, the absorption and 
scattering coefficients have been calculated differently
and with different optical constants, with a change in
amorphous carbon for the C stars and a change from 
astronomical silicates derived from observations to 
silicates measured in the laboratory for M stars.

The ratio of the MLRs for 76 non-foreground (FG) M stars 
(in the sense of old/new) has a median value of 1.17, with 
90\% of the ratios in the range 0.6 to 2.7.  The M stars 
show no obvious systematic effects, and the scatter suggests 
a random fitting error of a factor of two in the MLR.

\begin{figure}  
\begin{minipage}{0.49\textwidth}
\resizebox{\hsize}{!}{\includegraphics{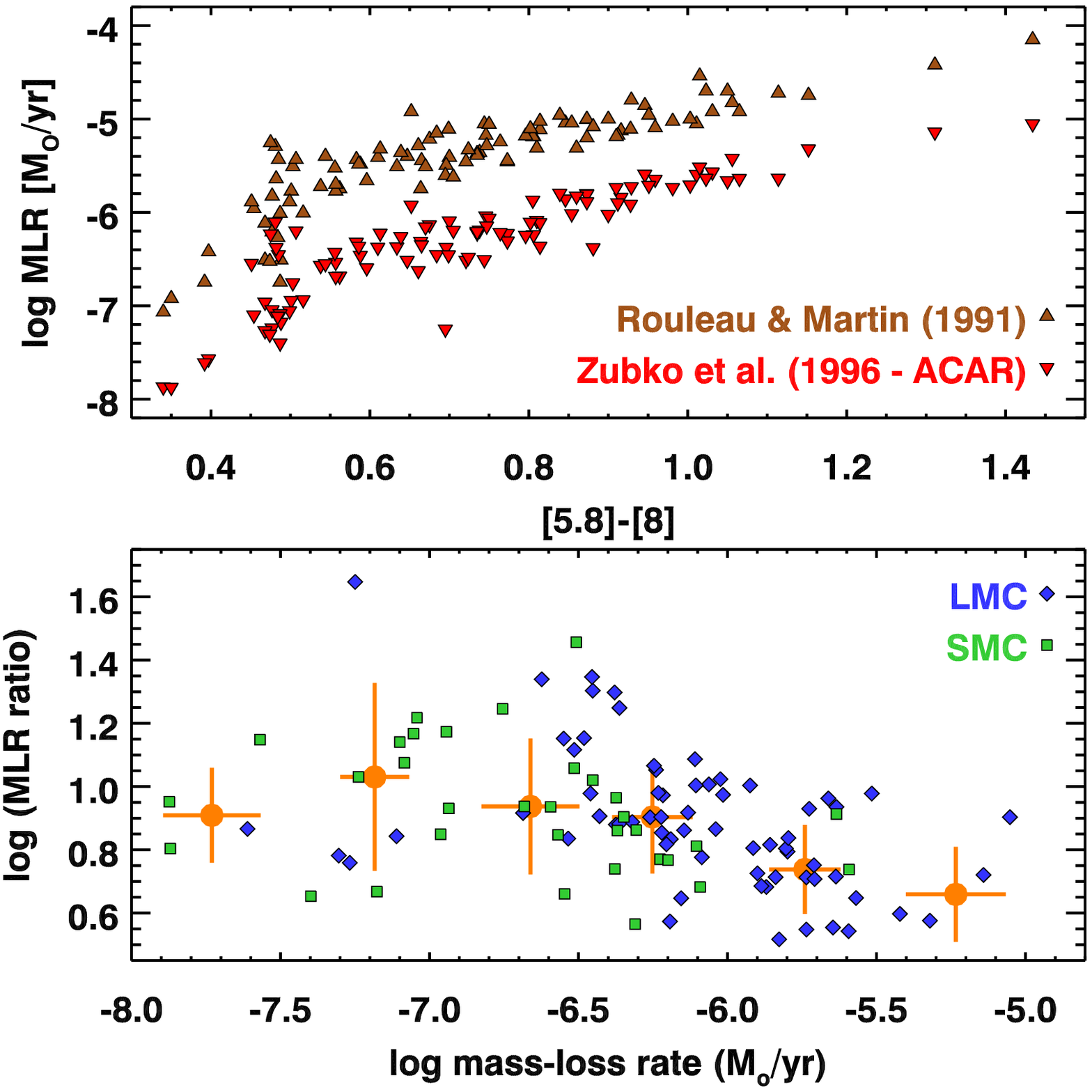}}
\end{minipage}
\caption[]{Comparison of the mass-loss rates for carbon 
stars between the present work and G09.  The current work
uses optical constants for amorphous carbon from Zubko et 
al.\ (1996), while G09 used constants from Rouleau \& Martin 
(1991).  Top:  A direct comparison for the common sources 
shows that the new models have much lower MLRs. 
Bottom:  The ratios of the MLRs tend to be approximately nine, although
the difference between the models drops for the highest MLRs.
The gold data points in the bottom panel are the median
values for the log of the MLR ratio in each 0.5-dex bin,
plotted against the mean log MLR; the error bars are the
formal standard deviations.}
\label{Fig-MLR} 
\end{figure} 

The difference between the current models and those of G09 
is much more dramatic for the carbon stars, as 
Figure~\ref{Fig-MLR} shows for the 101 sources in common 
between the two samples.  This difference arises primarily
from the shift from the optical constants for amorphous 
carbon from Rouleau \& Martin (1991) to the ACAR sample
from Zubko et al.\ (1996).  The mean ratio is 8.99, the
median is 7.58, and the standard deviation is 5.82.  80\% 
of the sample have a ratio of MLRs between 5.2 and 11.4.
The bottom panel of Figure~\ref{Fig-MLR} shows that the
change in MLR due to the change in optical constants 
decreases with higher MLR, down to a ratio of 5.0 for the
highest mass-loss bin.

A direct comparison of the extinction coefficient of the 
amorphous carbon grains used in the present work and by G09 
(they assumed small grains and ignored scattering) shows 
that the ratio of opacities is about 9.5 at 1~$\mu$m and 
5.4 at 2~$\mu$m, consistent with the trend seen in Fig.~\ref{Fig-MLR}.

These results reveal the impact of the chosen optical
constants for amorphous carbon.  The constants from Zubko
et al.\ (1996) produce slightly better fits to the observed
spectra, but no compelling reason exists for choosing one
set of constants over another.  The consequences of this
decision are substantial.  Boyer et al.\ (2012) and Matsuura 
et al.\ (2013) both estimated the total dust input from AGB
stars in the SMC and LMC, but their estimates differed by 
factors of approximately 8 in the LMC and $\sim$11 in the SMC.  In
both cases, the estimates by Matsuura et al.\ were higher,
because they used the models by G09 to calibrate the MLRs 
of their photometric sample, and those models were based
on the optical constants from Rouleau \& Martin (1991).
Boyer et al., on the other hand, used the GRAMS models
(Srinivasan et al.\ 2011), which are based on the ACAR
constants from Zubko et al.\ (1996).

Dust grains with more regular lattice structures should be
more efficient absorbers and emitters, because they can
function as better antennae.  If we can apply that principle
to the dust constants, it follows that more graphitic
carbon-rich dust will have higher opacities, and thus require
less dust mass to explain a given amount of emission and
absorption.  More amorphous dust would require more mass,
and that is consistent with the constants from Rouleau \&
Martin (1991), the models by G09, and the estimated total
dust inputs in the SMC and LMC by Matsuura et al.\ (2013).
On a microscopic level, graphitic grains can be described as
purely aromatic sheets of carbon, while more amorphous
grains would consist of mixtures of aromatic and aliphatic
carbon.  The aliphatic/aromatic ratio is thus the key
descriptor of the dust.  It would be particularly helpful
for future laboratory work to explore this ratio in their
samples, because we still do not have a way to distinguish
what best fits the carbon stars we have observed.  And as
noted already, this lack of knowledge leads to significant
uncertainty in total dust production by carbon stars in
nearby galaxies.

\subsubsection{Riebel et al.\ (2012)} 
\label{SS-Riebel}

\begin{figure}
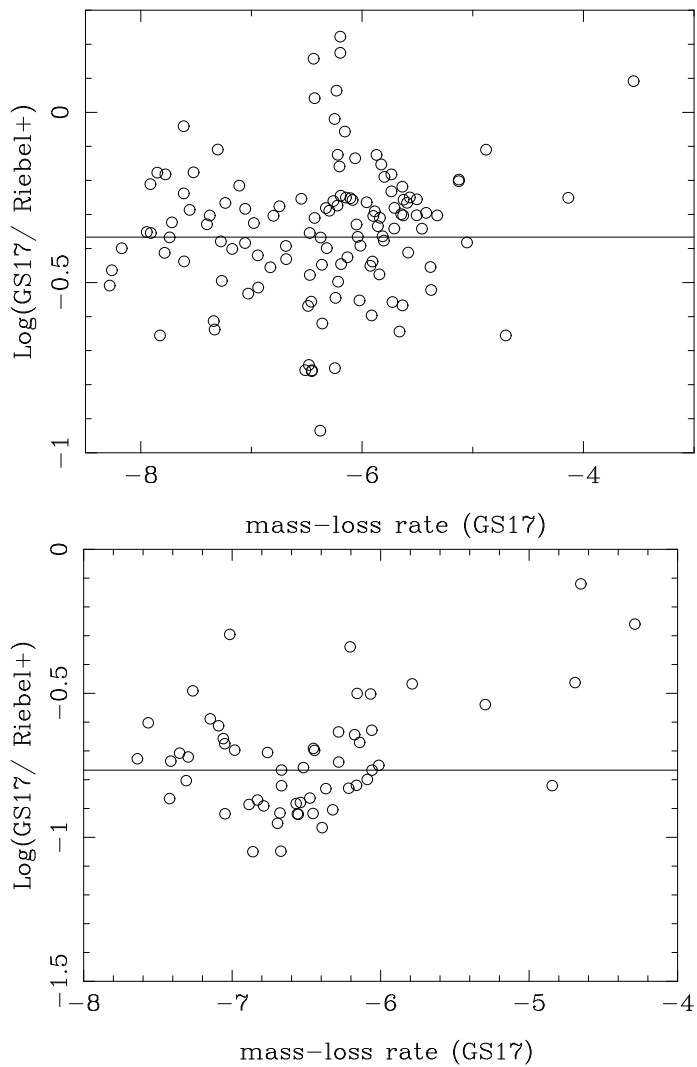
 

\begin{minipage}{0.49\textwidth}
\resizebox{\hsize}{!}{\includegraphics{compareRiebel_paper_C.ps}}
\end{minipage}
\begin{minipage}{0.49\textwidth}
\resizebox{\hsize}{!}{\includegraphics{compareRiebel_paper_O.ps}}
\end{minipage}

\caption[]{Ratio of dust MLRs found in the present paper 
and by Riebel et al.\ (2012).  In the top panel the C stars, 
in the bottom panel the O-rich stars.  The lines indicate the 
mean ratios of 0.46, and 0.17, respectively.}
\label{Fig-Riebel}
\end{figure}

We can also compare our results to those of Riebel et al.\ 
(2012), who derived MLRs for a sample of $\sim 30~000$ AGB 
stars and RSGs in the LMC by fitting up to 12 photometric 
bands with the GRAMS models.  We matched our source list to 
their LMC targets using a search radius of 1\arcsec\ and only 
keeping stars for which they list an error in the MLR of less 
than 30\%.  In Fig.~\ref{Fig-Riebel} we plot the ratio of our 
dust MLRs (found by dividing the MLR by 200) and theirs.  For 
130 C stars in common the median of this ratio is 0.46, with 
90\% of the stars within a factor of 2.7 of this.  For 63 M 
stars (excluding FG objects) the median ratio is 0.17 with 
94\% within a factor of three of this value.

The difference in MLRs could arise from different inner radii 
in case of the C stars (see next section) while for the 
O stars it could be due to the difference in using 
astronomical silicates versus opacities based on optical 
constants measured in the laboratory (see Sections~\ref{SS-J12} and \ref{SS-J14}).

\subsubsection{Srinivasan et al.\ (2011)} 
\label{SS-Sri11}

Srinivasan et al.\ (2011) compared their results to G09 and 
found that their MLRs were a factor 1.7 lower (median value).
They related this difference to their use of the optical
constants from Zubko et al.\ (1996), while G09 used the
constants from Rouleau \& Martin (1991).  However, as noted
in the previous section, the same change in optical constants
reduces the MLRs by a factor of between 5 and 10 for the
stars in common between the present work and G09.  Thus the
differences in methodology must also affect the estimated MLRs.

Table~\ref{Tab-TRM} lists the four stars used to calibrate 
the GRAMS models.  They are the C stars TRM~88 and OGLE 
LMC-LPV-28579 (our identifier is ogle051306), and the M stars
HV~5715 and SSTISAGE1C J052206.92-715017.6 (our identifier is 
sagemcj052206).  The table lists the derived luminosities and 
dust MLRs in the various papers, with the error or range in 
the parameters when available.

Sargent et al.\ (2010) and Srinivasan et al.\ (2010)
describe the detailed radiative transfer modelling of the SED 
and IRS spectrum of the two M stars and C stars, 
respectively.  The same numerical code, optical constants, 
grain-size distribution, etc., derived in these papers were
then used in the generation and validation of the GRAMS model 
grid (Sargent et al.\ 2011, Srinivasan et al.\ 2011), and its 
application (Riebel et al.\ 2012, Jones et al.\ 2012, 
Srinivasan et al.\ 2016).  Other papers have modelled the SED 
and/or IRS spectrum using independent methods (e.g.\ van Loon 
et al.\ 1999, G09, Jones et al.\ 2014).  

It is important to note that the estimates for luminosity and 
MLR by Srinivasan et al.\ (2011, 2016), Sargent et al. (2011) 
and Riebel et al.\ (2012) are based on the same GRAMS model 
grid.  These efforts differed in the details of how the 
photometric data were gathered and the models were fitted, 
but they all used models from the same grid.  Their estimated 
luminosities and dust MLRs agree well, although, as discussed 
below, the work of Jones et al.\ (2012), which also used the 
GRAMS model grid, differed more significantly.

Section~\ref{SS-Riebel} quoted median ratios for our dust MLR 
to those by Riebel et al.\ of 0.46 for C stars and 0.17 for O 
stars.  Within the errors these ratios are consistent with 
the values for the individual objects (0.42 and 1.65, 
respectively, for the C stars TRM~88 and ogle051306, and 0.16 
and 0.21 for the two O stars).

However, the present work determines a dust MLR for TRM~88 approximetely eight 
times lower than G09.  This difference cannot be due
entirely to the different optical constants.  Other 
differences in the approach by us, G09, and Riebel et al.\
(2012) must also play a role.  

A likely suspect is that the GRAMS models allow for larger 
temperatures at the inner radius.  The GRAMS models are run 
for a fixed grid of inner radii ($R_{\rm in}$= 1.5, 3, 4.5, 
7, 12~\rstar\ for the C-star grid, and 3, 7, 11, and 
15~\rstar\ for  O-star grid), but only models with 
corresponding temperatures below 1800~K and 1400~K 
(respectively) are kept.  The present work does not accept
condensation temperatures higher than 1200~K.  Srinivasan et 
al.\ (2011) provide $R_{\rm in}$ or $T_{\rm c}$ for our
calibration C stars.  For TRM~88, we find an inner radius
of 15.6~\rstar, while Srinivasan et al.\ find a lower
value, 12~\rstar, which is consistent with the ratio
of dust MLR of 0.42 between the present work and the GRAMS
grid.  For OGLE~051306 we find 14.1~\rstar, while 
Srinivasan et al. (2010) find 4.4~\rstar, a difference
due to the temperature at the inner radius, 970 versus 
1300~K.  If we had adopted that temperature, our MLR would 
drop by factor of 3.2, and the ratio of our dust MLR compared 
to Riebel et al. (2012) would decrease from 1.65 to 0.5, or 
close to the median value.

\subsubsection{Jones et al.\ (2012)} 
\label{SS-J12}

For the O-rich stars HV~5715 and sagemcj052206, Sargent et 
al.\ (2010) fitted the SED and IRS spectra, and their results 
agree well with the GRAMS-based results of Sargent et al.\ 
(2011) and Riebel et al.\ (2012) (see Table~\ref{Tab-TRM}). 
The fitting method of Jones et al.\ (2012), however, led to 
a much higher estimate of dust MLRs.  They also used 
the GRAMS model grid, but the details of their method 
differed\footnote{They used an extra data point in the
SED corresponding to the IRAS 12 $\mu$m band and determined
from the IRS data, used larger error bars in the $\chi^2$ 
fitting in order to account for variability, resulting in 
more GRAMS models providing `good' fits to the data.  They 
also accounted for inclination angle of the LMC disk which 
leads to some differences in luminosity and MLR (Jones et 
al., 2012, private communication).}.

An examination of the 69 M stars in common between Jones et 
al.\ (2012) and Riebel et al.\ (2012) leads to a median 
ratio in the MLRs of 1.6 (Jones et al.\ / Riebel et al.), 
which is not large, but only 35\% of the stars lie within a 
factor of three of the median.  Nineteen stars have MLR ratios 
which differ from the median by a factor of 10 or more, and 
five differ by a factor of 75 or more. Thus the scatter when 
considering individual objects is significantly higher, even 
if the statistical difference for the overall sample is 
small.  As noted above, for 63 M stars in common between 
Riebel et al.\ (2012) and the present work, the median ratio 
(this work / Riebel et al.) in the MLRs is 0.17, with no object 
with an MLR ratio more than a factor of ten from the median. 
That difference presents the opposite problem: a reasonable 
consistency among source-to-source, but a greater shift between 
the model results overall.

\subsubsection{Jones et al. (2014) } 
\label{SS-J14}

Jones et al.\ (2014) investigated a sample of evolved stars 
in the LMC by fitting photometry and IRS spectra to a grid of 
models calculated with the code {\it MODUST} (Bouwman et al.\ 
2000).  Table~\ref{Tab-TRM} shows that for HV~5715 and 
sagemcj052206, they find much lower dust MLR than Jones et 
al.\ (2012), and this result is generally true for the larger
sample.  The two samples contain 26 stars in common, and the 
median ratio of the 2014 results to 2012 is 0.13, again with 
a large scatter; for 30\% of the stars the difference exceeds
a factor of five.  The major difference between the two works is 
the adopted opacities:  Jones et al.\ (2012) relied on the 
GRAMS models which use the `astronomical silicates' from 
Ossenkopf et al.\ (1992), while Jones et al.\ (2014) derive 
the opacities from optical constants measured in the 
laboratory, as in the present work.

The differences between the dust MLR from Jones et al.\ 
(2014) and the present work are relatively small and 
uniform.  For 18 stars in common, the median ratio is 1.65, 
with all stars within a factor of 2.6 of this value.  
Because both papers used identical optical constants for 
amorphous silicates and aluminium oxide, the differences are
likely due to the differences in the opacity for iron and 
the derived (present paper) or adopted (Jones et al.\ 2014) 
iron fraction.  Differences in the radiative transfer and
fitting procedure are likely to have had a smaller effect.
We typically find larger iron fractions than adopted by Jones 
et al.\ (2014) and hence lower MLRs.

We have calculated the extinction coefficient for warm 
oxygen-deficient silicates from the astronomical silicates
from Ossenkopf et al.\ (1992) for single-sized grains of 
0.15~\mum\ and compared those to the grains that best fit 
HV~5715 and sagemcj052206 in the present work.  The grains
in the present work are larger, and the ratio of opacities at 
1 and 2~\mum\ are 1.5--2.6 and 2--6, respectively, consistent 
with the differences in MLRs between the present work and 
most of the works based on the M-star GRAMS grid.

\begin{table}

  \caption{Comparison of derived dust MLRs and luminosities 
for two C stars (TRM~88 and OGLE~051306) and two O stars.}
  \begin{tabular}{lrrrrrrr}
  \hline
   Reference   &  $L$      & \mdot$_{\rm dust}$  \\
               &  (\lsol)  & (10$^{-9}$ \msolyr) \\
\hline
\multicolumn{3}{c}{\underline{ TRM~88} }  \\ 
van Loon et al. (1999)   & 13300           &  6.0 \\ 
G09                      &  7160           & 16.1 \\ 
Srinivasan et al. (2011) & 11700           &  3.4             \\ 
Riebel et al. (2012)     &  9120 $\pm$ 650 &  4.92 $\pm$ 0.58 \\ 
present paper            &  9403           &  2.1             \\ 

& \\
\multicolumn{3}{c}{\underline{ OGLE~051306 } }  \\ 
Srinivasan et al. (2010) &  4810, 6580     &  2.5 (2.4-2.9)   \\ 
Srinivasan et al. (2011) &  6170           &  2.4             \\ 
Riebel et al. (2012)     &  7080 $\pm$ 700 &  2.12 $\pm$ 0.42 \\ 
present paper            &  4740           &  3.2             \\ 

& \\
\multicolumn{3}{c}{\underline{ HV~5715} }  \\ 
Sargent et al. (2010) & 36~000 $\pm$ 4000 &  2.3 (1.1-4.1)   \\ 
Sargent et al. (2011) & 33~000            &  1.5             \\
Riebel et al. (2012)  & 33~700 $\pm$ 5960 &  1.56 $\pm$ 0.43 \\ 
Jones et al. (2012)   & 28~800            & 19.6             \\ 
Jones et al. (2014)   & 19~230 $\pm$ 4300 &  0.63 $\pm$ 0.14 \\ 
present work          & 28~200            &  0.25            \\ 

& \\
\multicolumn{3}{c}{\underline{ sagemcj052206} }  \\ 
Sargent et al. (2010) & 5100 $\pm$ 500 &  2.0 (1.1-3.1)   \\ 
Sargent et al. (2011) & 4900           &  2.1             \\ %
Riebel et al. (2012)  & 4820 $\pm$ 670 &  2.11 $\pm$ 0.44 \\ 
Jones et al. (2012)   & 4740           & 19.6             \\ 
Jones et al. (2014)   & 3160 $\pm$ 710 &  0.68 $\pm$ 0.19 \\ 
present work          & 4120           &  0.45            \\ 

\hline 
\end{tabular} 

\label{Tab-TRM}
\end{table}

\subsection{Mass-loss and stellar evolution} 
\label{sec-ML}

G09 compared the MLR vs.\ luminosity diagram with predictions 
from stellar evolutionary models by Vassiliadis \& Wood 
(1993, hereafter VW) and those using recipes developed by
Wagenhuber \& Groenewegen (1998) with a Reimers mass-loss law 
(with multiplicative scaling parameter $\eta = 5$ on the 
AGB).  The comparison largely favoured the VW models, 
vindicating their adopted mass-loss recipe, which is 
essentially the minimum of the single scattering limit 
\mdot $= 2.02 \cdot 10^{-8} \, L / v_{\rm exp}$, and an 
empirical relation between $\log \dot{M}$ and $P$.  The 
fundamental-mode period, $P$, is calculated from a 
period-mass-radius relation (see VW for details).

\begin{figure} 

\begin{minipage}{0.49\textwidth}
\resizebox{\hsize}{!}{\includegraphics{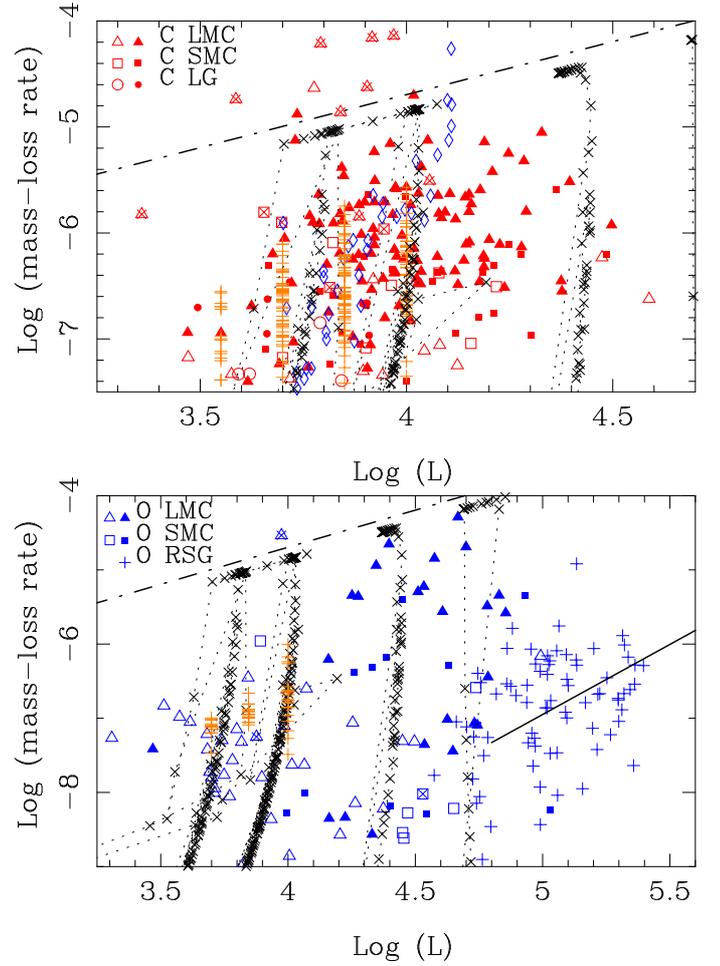}}
\end{minipage}

\caption[]{Total MLR versus luminosity for C stars (top 
panel, red colours) and M stars (bottom panel, blue colours).
Objects with Mira-like pulsation amplitudes are plotted with 
filled symbols; objects with smaller amplitudes with open 
symbols.  An overplotted cross indicates no information on 
pulsation properties.  
Orange plus signs in the top panel indicate models by 
Eriksson et al.\ (2014) scaled to our adopted dust-to-gas 
ratio and expansion velocity (see text).  Blue diamonds in 
the top panel indicate a sample of Galactic C Miras (see 
text).  RSGs are plotted as plus signs independent of host 
galaxy and pulsation amplitude.  
Orange plus signs in the bottom panel indicate models 
by Bladh et al.\ (2015) scaled to our adopted dust-to-gas 
ratio and expansion velocity (see text).
The VW models for LMC 
metallicity are plotted as crosses connected by the dotted 
lines for initial masses of 1.5, 2.5, 5.0 and 7.9~\msol, but 
not every track is visible in every panel. Each cross 
represents a time interval of 5000 years.  The dot-dashed 
line shows the single scattering limit for a velocity of 
10~\ks.  The solid line is the relation found by Verhoelst 
et al.\ (2009) for Galactic RSGs.
} 
\label{Fig-MLLum} 
\end{figure} 

\begin{figure} 

\begin{minipage}{0.49\textwidth}
\resizebox{\hsize}{!}{\includegraphics{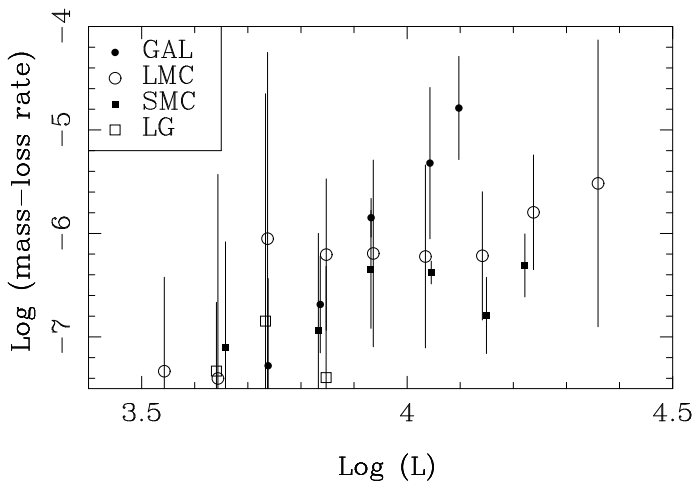}}
\end{minipage}

\caption[]{ 
Binned version of Fig.~\ref{Fig-MLLum} for C stars.
}
\label{Fig-binned}
\end{figure} 

Figure~\ref{Fig-MLLum} shows the relation between MLR and 
luminosity, with the VW model tracks for LMC metallicity 
overplotted (the crosses connected with the dotted lines). 
From the individual evolutionary tracks, a model is plotted 
every 5000 years.  The density of points therefore represents
the time spent at a certain position in the diagram.  It also 
explains the `excursions' which are due to the finite 
probability of catching a star during a thermal pulse or 
the luminosity dip that follows.  Models are plotted for 
initial masses of 1.5, 2.5, 5.0 and 7.9~\msol, which evolve 
at increasing luminosity. 

The distribution of MLRs for the O-rich AGB stars (in the
bottom panel) is similar to that described by G09.  They are 
prominent at lower luminosities, which correspond to lower 
masses where C stars are less likely to form or will form 
later during their evolution on the AGB.  Oxygen-rich AGB 
stars are largely absent at intermediate luminosities, which 
correspond to masses dominated by C stars, and they are 
prominent at higher masses, where hot-bottom burning inhibits 
the formation of C stars. Essentially no stars exceed the
single-scattering limit. The MLRs of the RSGs scatter around 
the relation for Galactic RSGs derived by Verhoelst et al.\ 
(2009), $\log \dot{M}= -16.4 + 1.89 \log L/L_{\odot}$.
The spread is $\sim$2 dex, which is larger than the 
$\sim$1-1.5 dex scatter in the relation for Galactic RSGs.

The orange plus signs represent the recent dynamical 
models by Bladh et al.\ (2015) for M stars with masses 
of 1~\msol\ and solar metallicities.  The models are 
available for 5000, 7000, and 10000~\lsol\ and cover a range 
in effective temperatures, piston-velocity amplitudes and 
seed-particle densities.\footnote{The models with 
seed-particle density $10^{-16}$ were not considered, as they 
do not fit the data considered by Bladh et al.\ very well 
(Bladh 2017, private communication).}
The plotted MLR is their value scaled to our adopted
dust-to-gas ratio and expansion velocity.  The differences
are significant.  The median values in the 56 models of 
dust-to-gas ratio ($\Psi$) and $v_{\rm exp}$ are 5.8 $\cdot 
10^{-4}$ and 10.4~\ks, respectively.  The difference in 
$\Psi$ is considerable, and their calculated MLRs 
are a median 1.1 dex above the scaled values plotted in 
Figure~\ref{Fig-MLLum}.
A comparison to our determinations is difficult because of 
the difference in metallicity and mass, but the agreement is
within an order of magnitude, which is encouraging given the 
fact that the driving of winds in M stars is a difficult 
problem (Woitke 2006, H\"ofner 2008).

The top panel of Figure~\ref{Fig-MLLum} shows the results for 
the C stars.  The orange plus signs are the results of the 
dynamical models at solar metallicity by Eriksson et al.\ 
(2014).  From the 540 models they calculated we show the 193 
models which result in a outflow (not all models do), and 
then those with an expansion velocity larger than 5 \ks.  The 
plotted MLR is their value scaled to our adopted dust-to-gas 
ratio and expansion velocity.  This difference is significant. 
The median values for the 193 models of dust-to-gas 
ratio ($\Psi$) and $v_{\rm exp}$ are 0.0014 and 23.4 \ks, 
respectively, which differ considerably from our values.  
Their calculated MLRs lie a median 0.91 dex above 
the scaled values plotted in Figure~\ref{Fig-MLLum}, and 
reach and exceed the single-scattering limit (see below).

The different models at a given luminosity are related to 
differences in the other input parameters like effective 
temperature, overabundance of carbon, or velocity of the 
piston at the inner boundary of their models.

Figure~\ref{Fig-MLLum} also includes Galactic stars (plotted
as blue diamonds).  The fitting for the Galactic targets
was done using different software and dust opacities, which
we can expect to lead to differences in derived MLRs (see
Sect.~\ref{sect-comp}).  This sample includes three groups.

Groenewegen et al.\ (1998) modelled the SEDs and spectra of
42 Galactic carbon-rich Miras.  In this case, the spectra
were from the Low-Resolution Spectrometer (LRS) aboard the
{\it Infrared Astronomical Satellite} ({\it IRAS}).  Miras
were chosen so that the period-luminosity relation could be
used to estimate their distances, which are always a
challenge for evolved stars in the Galaxy.  As part of a
different (unpublished) investigation, six stars in that
sample covering a range of MLRs were fitted with MoD to
derive the differences in MLRs, using updated photometry and
where possible, spectra from the Short-Wavelength
Spectrometer (SWS) aboard the {\it Infrared Space 
Observatory} ({\it ISO}).  The third group comes from 
Groenewegen et al.\ (2016) who modelled some very red 
Galactic objects with MoD using the same method as in 
the present paper.  That sample included three additional 
stars overlapping Groenewegen et al.\ (1998), and two other 
stars.  Thus the Galactic sample consists of 44 stars:  33 
taken directly from the models by Groenewegen et al.\ (1998), 
nine from that sample based on updated models, and two 
additional stars.

The median ratio of the MLRs (old/new) is 3.8, and this is
largely due to the change in optical constants (Groenewegen
et al.\ 1998 also used the constants from Rouleau \& Martin 
1991).  We have scaled the 33 old models by this median 
ratio.  We have also scaled the MLRs of these stars to an 
expansion velocity of 10~\ks, as adopted in the rest of the 
sample.  The stars follow a close relation which is due to 
the underlying adopted period-luminosity relation.

Compared to G09, the qualitative description of the 
comparison has changed.  G09 found that only three C stars 
were slightly above the single-scattering limit, which they 
considered to be consistent with expectations.\footnote{The single-scattering limit applies when $\beta = 1$, where 
$\beta \equiv$ (\mdot $v_{\rm exp})/(L/c)$, the ratio of the 
matter-momentum flux to the photon-momentum flux.}  In the 
current picture, a significant number of C stars are above 
the single-scattering limit by up to a factor of ten.
This in itself does not necessarily pose a significant issue. 
First, the MLRs have large uncertainties (and $\beta$, too, 
due to the assumed gas-to-dust ratio of 200 and expansion 
velocity of 10 \ks). 
Second, dynamical models show that $\beta >$ 1 can be reached 
in realistic models. 
From a subset of 900 dynamical model atmospheres for carbon 
stars for solar metallicities from Mattsson et al.\ (2010), we 
find that the 98th percentile on $\beta$ is 3.0.  In the more 
recent models by Eriksson et al.\ (2014), the 98th percentile 
on $\beta$ is 1.5.  From dynamical model atmospheres for 
subsolar metallicities by Wachter et al.\ (2008) one might 
expect the value for $\beta$ to be a factor 2.6 lower in the 
LMC, i.e.\ near 0.5-1.

If confirmed, our models show that the artificial cut-off in 
the VW models at $\beta = 1$ may be too conservative.  A 
cut-off (if any) at a larger $\beta$ would result in shorter 
AGB lifetimes.

Figure~\ref{Fig-binned} presents the results differently, 
with the MLRs of the C stars binned and averaged for  
the SMC, LMC, dSph galaxies, and our Galaxy separately.  The 
VW evolutionary tracks suggest that for a given mass, the 
luminosity evolves by about 0.1 dex on the AGB 
(Fig.~\ref{Fig-MLLum}), and as a consequence we chose this as 
the bin size in $\log L$.  The MLRs in a luminosity bin are 
median averaged in $log$ \mdot\ and are plotted in 
Fig.~\ref{Fig-binned} at the average luminosity of the stars 
in that bin if a bin includes three or more objects.  The 
error bar indicates the spread in the bins (as 1.5 times the 
median absolute deviation).

The MLR increases globally with luminosity, as also 
shown by the models by Eriksson et al.\ (2014).  Any 
dependence on metallicity remains difficult to assess.  The 
issue of accurate distances (hence luminosities) remains a 
limiting factor for any Galactic sample.  The models in the 
present work point to a larger dust MLR in the LMC than in 
the SMC for a given luminosity, but this could also 
arise from the difference in the underlying populations (see 
below) and/or differences in expansion velocity.

Groenewegen et al.\ (2016) determined for the first time the 
expansion velocity of four C stars in the LMC by detecting the 
CO J=2-1 transition using the Atacama Large-Millimeter Array
(ALMA).   All four of these stars are in the present sample:
IRAS~05506, IRAS~05125, ERO~0529379, and ERO~0518117. 
They compared these objects to the closest 
available analogs in the Galaxy and found that the expansion 
velocity in the LMC appears to be smaller than in the Galaxy.  
The key caveat, though, is that the samples are small, and it 
is difficult to find suitable comparison objects in the two galaxies.

Figure~\ref{Fig-MLLum} shows clearly that between the LMC
and SMC, the stars with the heaviest mass loss are located in 
the LMC.  This result does not appear to result from a bias
in the spectroscopic sample observed by the IRS.  Plotting 
colour-magnitude diagrams (CMDs) of photometric samples in the
mid-IR reveals more intrinsically red AGB stars in the LMC
compared to the SMC (Fig.~5 from Ventura et al.\ 2016, and 
references therein).  Comparison to models shows that the 
largest degree of obscuration in the LMC and SMC occurs for
stars with an initial mass of 2-3~\msol\ and about 1.5~\msol, 
respectively, a difference which Ventura et al.\ (2016)
attribute to differences in the star formation 
histories of the two galaxies.  Such differences between the 
populations in the LMC and SMC make it difficult to draw any 
firm conclusions about how the MLR depends on metallicity.

\subsection{A super-AGB star candidate} 
\label{sec-SAGB}

G09 suggested that MSX SMC 055 (or IRAS 00483$-$7347) is a 
good candidate for a super-AGB (SAGB) star, based on its high 
luminosity ($M_{\rm bol}= -8.0$), its very long 
pulsation period (1749 days) and large pulsation amplitude 
(1.6~mag peak-to-peak in $I$).  Its pulsational properties 
distinguish it from a luminous RSG.
Here we present improved estimates for its parameters.

Groenewegen \& Jurkovic (2017) derived a relation between 
period, luminosity, mass, temperature, and metallicity based 
on the 5-11~\msol\ initial-mass Cepheid models by Bono et al.\ (2000).
For the parameters $P$= 1810 $\pm$ 50 days, $L$= 85350 $\pm$ 
8500~\lsol, $T_{\rm eff}$= 2500 $\pm$ 100~K, and $Z$= 0.004 
$\pm$ 0.001 (errors are adopted), we derive a current 
pulsation mass of 8.5 $\pm$ 1.6~\msol\ with the total error 
in mass dominated by the error in effective temperature.
The simpler period-mass-radius relation for fundamental-mode 
Mira pulsators from Wood (1990) gives a similar value of 9.2 
$\pm$ 1.8~\msol.

The current MLR is estimated to be 4.5 $\cdot 10^{-6}$ \msolyr, 
assuming a conservative gas-to-dust ratio of 200.
Roman-Duval et al.\ (2014) estimate a value in the ISM in 
the SMC of 1200$^{+1600}_{-420}$ which implies the MLR could 
be larger by a factor of a few.
Lifetimes in the thermal-pulsing phase are short in SAGB 
stars ($10^{4-5}$ years; e.g.\ the review by Doherty et al.\ 2017), 
but these lifetimes in combination with a MLR that 
could exceed $10^{-5}$ \msolyr\ indicate that the 
initial mass of MSX SMC 055 could very well be up to 1~\msol\ 
larger than its current mass.

Garc\'ia-Hern\'andez et al.\ (2009) observed this and other 
massive AGB star candidates in the MCs in the optical at high 
spectral resolution.  The source MSX SMC 055 is very rich in 
rubidium, Rb, with [Rb/Z] \more  +1.7, which confirms the 
activation of the $^{22}$Ne neutron source at the $s$-process 
site and its massive AGB or SAGB nature.  Indeed, by 
comparing these observations to models, they independently 
suggested an initial mass of at least 6--7~\msol\ for this star. 

The most viable SAGB star candidate in the SMC and LMC remains MSX~SMC~055.

\subsection{The potential of JWST} 
\label{sec-JWST}

The James Webb Space Telescope (JWST), due for launch in
2019, provides impressive filter sets on its two
imaging instruments, NIRCAM and MIRI, which will enable
broadband photometric studies of the evolved stellar
population in galaxies out to distances of a few Mpc.  At
least two published papers have already investigated which
filter combinations look to be most useful for distinguishing
and characterizing different classes of objects.

Kraemer et al.\ (2017) examined the sample of SMC objects
observed by the IRS, using the spectra to confirm the
classifications.  They found that the 5.6, 7.7, and 21~\mum\
filters best separated C-rich from O-rich stars, while the
5.6, 10, and 21~\mum\ filters best separated young stellar
objects (YSOs) from planetary nebulae (PNe).  Jones et al.\
(2017a) performed a similar study using over 1000 sources
with IRS spectra in the LMC.  They discussed how to
discriminate O- and C-rich AGB and post-AGB stars, RSG,
H{\sc ii} regions, PNe and YSOs.  Both Kraemer et al.\ and
Jones et al.\ based their CMDs and CCDs on synthetic
photometry using the IRS spectra.  Therefore, they were
limited in showing diagrams based on MIRI filters.  In
Appendix~\ref{App-C} we present synthetic photometry for the
sample of almost 400 evolved stars in about 75 filters,
including the 29 medium and wide-band filters available with
the NIRCAM and MIRI instruments.

\begin{figure}
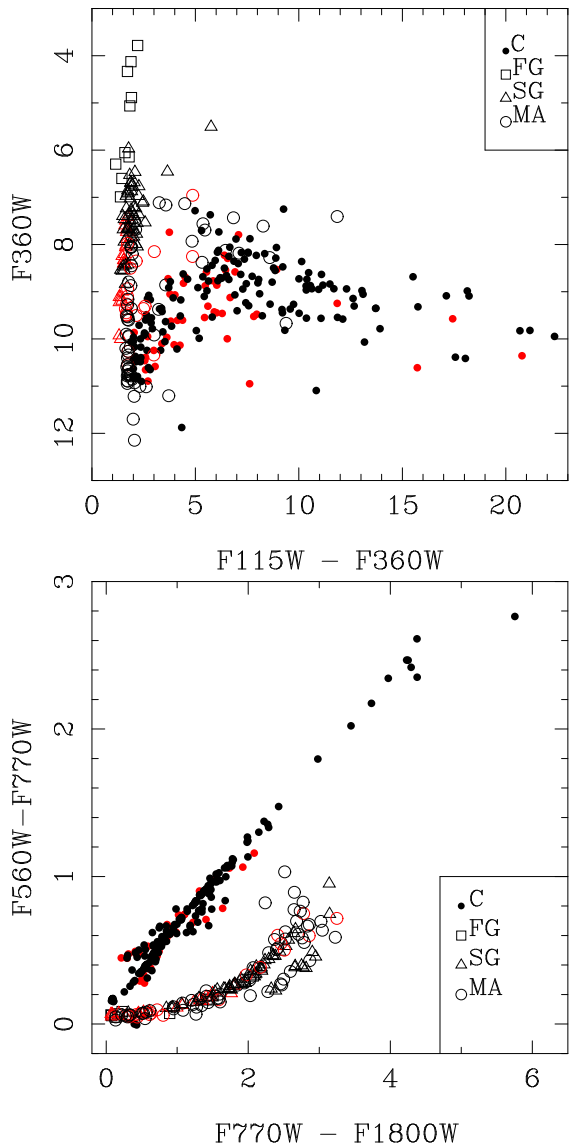


\begin{minipage}{0.4\textwidth}
\resizebox{\hsize}{!}{\includegraphics{jwst_12_360_115.ps}}
\end{minipage}

\begin{minipage}{0.4\textwidth}
\resizebox{\hsize}{!}{\includegraphics{jwst_3_56_77_180.ps}}
\end{minipage}

\caption[]{Colour-Magnitude and Colour-Colour diagrams for the 
LMC (with the SMC stars, plotted in red, shifted to the distance 
of the LMC) based on synthetic JWST NIRCAM/MIRI magnitudes.
}
\label{Fig-HRD} 
\end{figure}

Figure~\ref{Fig-HRD} gives two examples.  The first is a CMD
resembling the near-infrared and IRAC [3.6] versus $J$$-$[3.6]
diagram shown by G09 (SMC objects have been placed at the
distance of the LMC).  In this diagram, the dustiest AGB
stars show a relatively small spread in F360W magnitude and
the reddest objects are predominantly C-rich.  The second
example is a CCD resembling the [5.8]$-$[8.0] versus
[8.0]$-$[24] diagram shown in Fig.~\ref{Fig-CCD} which
separates O-rich and C-rich very well.  Replacing the F1800W
filter by F2100W or F2550W yields simililar plots, but for a
given integration time and signal-to-noise F1800W is a
magnitude more sensitive than F2100W and almost three magnitudes
more sensitive than F2550W (Jones et al.\ 2017a).
Thus for AGB stars, we recommend the 5.6, 7.7, and 18~\mum\
filters to discriminate O-rich from C-rich stars.

\section{Summary and conclusions} 

We have fitted the SEDs and {\it IRS} spectra of almost 400
evolved stars in the SMC and LMC with a dust
radiative-transfer model to determine luminosities and (dust)
mass-loss rates.  The mass-loss rates depend strongly on the
adopted opacities (that is, the optical constants, and 
to a lesser extent the grain shape).

A comparison with results in the literature shows that for M 
stars the choice of optical constants based on laboratory 
measurements leads to lower MLRs than those derived from 
observations (so called, astronomical silicates) as employed 
in the widely used GRAMS models.  When using 
laboratory-determined optical constants, the iron content 
that is assumed or derived becomes important and introduces an uncertainty 
of a factor of two.

For C stars the choice between the widely used optical 
constants by Rouleau \& Martin (1991) and Zubko et al.\ 
(1996) introduces a difference in MLR of a factor of 
approximately five or more.  Comparison with the literature suggests that 
differences in the allowed inner radius in the radiative 
transfer modelling may also introduce a factor of two 
uncertainty in the derived MLR.

All of these uncertainties impact the estimates of the total
gas and dust return of evolved stars in the MCs (see the
references in Section~\ref{sec-mlr}).  The differences in
opacities are the greater problem, and the solution lies in a
better determination of what circumstellar grains actually
look like.  While this paper does not offer any solutions to
the problem, we hope that we have helped to better frame the
problem and its consequences on our understanding of the role
played by AGB stars in galactic evolution.

Also of interest are the particular cases of the half-dozen
sources with the largest optical depths.  They are not the
most luminous sources.  Groenewegen et al.\ (2016) and
Sloan et al.\ (2016) discussed their evolutionary status.
Evolutionary models using the \texttt{COLIBRI} formalism
described by Marigo et al.\ (2013) (see additional detail by
Groenewegen et al.\ 2016) agree with Ventura et al.\ (2016)
that these stars began their lives with initial masses of
2--3~\msol, but have had their envelope masses reduced to
$\less$1~\msol\ through the mass-loss process.  The low
envelope mass is necessary to explain their long pulsation
periods (often longer than 1000 days).  Not all of the
reddest stars have had their pulsation properties determined,
which would clearly be an important contribution.  Sloan et
al.\ (2016) noted that the reddest sources showed decreased
variability and evidence that radiation from the central
star may be escaping the otherwise optically thick dust
shell.  Both behaviours are consistent with evolution off of
the AGB.  The unusual blue colours suggest that these stars
may also be departing from spherical symmetry, in which case
our radiative-transfer models could be underestimating their
luminosity.  These may be the sources at the very end of
their AGB lifetimes, and we need better observational
constraints on their geometry and outflows.

\begin{acknowledgements}
We thank the anonymous referee for a very thorough report,
which led to numerous improvements in this manuscript.
GCS was supported by NASA through Contract Number 1257184
issued by the Jet Propulsion Laboratory, California Institute 
of Technology under NASA contract 1407 and the NSF through 
Award 1108645.  
We thank Jean-Baptiste Marquette (IAP) for providing some 
EROS2 photometric data, Alistair Glasse (UKATC) for providing 
the JWST/MIRI filter response curves, and Sara Bladh (Padova 
University) for providing the results for her M-star models, 
including the gas-to-dust ratios.
This publication is partly based on the OGLE observations 
obtained with the Warsaw Telescope at the Las Campanas 
Observatory, Chile, operated by the Carnegie Institution of 
Washington.
This paper utilizes public domain data originally obtained by 
the MACHO Project, whose work was performed under the joint 
auspices of the U.S. Department of Energy, National Nuclear 
Security Administration by the University of California, 
Lawrence Livermore National Laboratory under contract No.\
W-7405-Eng-48, the National Science Foundation through the 
Center for Particle Astrophysics of the University of 
California under cooperative agreement AST-8809616, and the 
Mount Stromlo and Siding Spring Observatory, part of the
Australian National University.
This research made use of the SIMBAD database, operated at 
CDS, Strasbourg, France. 
This publication makes use of data products from {\it WISE}
and {\it NEOWISE}, which a joint projects of the University 
of California, Los Angeles, and the Jet Propulsion 
Laboratory/California Institute of Technology, funded by the 
National Aeronautics and Space Administration.
Based on data from the OMC Archive at CAB (INTA-CSIC), 
pre-processed by ISDC.
The CSS survey is funded by the National Aeronautics and 
Space Administration under Grant No. NNG05GF22G issued 
through the Science Mission Directorate Near-Earth Objects 
Observations Program.  The CRTS survey is supported by the 
U.S.~National Science Foundation under grants AST-0909182.
\end{acknowledgements}

\makeatletter
\renewcommand\@biblabel[1]{}
\makeatother

{}

\begin{appendix}

\section{The sample}
\label{App-A}

Tables~\ref{Tab-Csample} and \ref{Tab-Msample} list the 
samples of C-rich and O-rich stars, with basic information:
some common names (as listed by SIMBAD), R.A. and 
declination in decimal degrees, the identifier used in 
figures and tables below (where we almost always kept the 
target name used in the original observation), the adopted 
pulsation period, the (semi-)amplitude and in which filter.
For the oxygen-rich stars a classifier is added 
(FG=Foreground, SG=Supergiant, MA=M-type AGB-star).
These tables are published in its entirety at the CDS; a portion is shown here for guidance regarding its form and content.

The C-stars are listed in order of R.A., but with the 
stars in the dSph galaxies listed last.  For the O-rich stars 
the foreground objects are listed before the SGs and the AGB 
stars, and then by R.A.
Various criteria have been used to 
distinguish between foreground, supergiant and O-rich AGB 
stars; see G09 for details.  Compared to G09 only one 
additional FG object was added HD~269924, which is a K5 star 
with a significant proper motion.  The distinction between 
SG and MA is sometimes not clear-cut, but our conclusions 
do not depend on the misclassification of a few objects.

\begin{figure}

\begin{minipage}{0.49\textwidth}
\resizebox{\hsize}{!}{\includegraphics{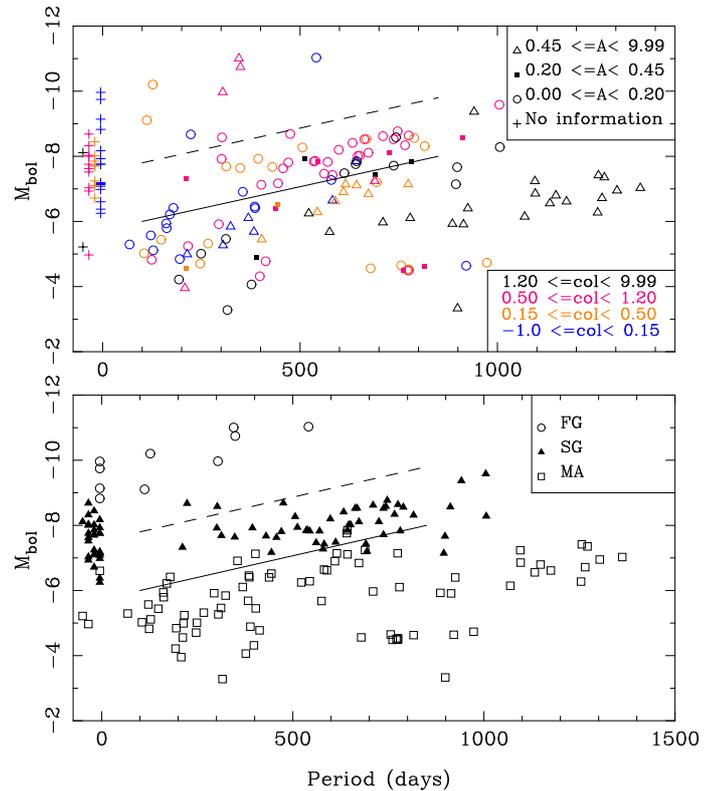}}
\end{minipage}

\caption[]{
Bolometric magnitude versus pulsation period for the O-rich stars.
Stars without periods are plotted as plus signs at negative periods.
The solid line indicates the lower luminosity limit for 
RSGs by WBF, and the dashed line is 1.8~mag brighter.
{\em Top panel}.
The legend indicates the meaning of the symbols in terms of 
the $I$-band semi-amplitudes ($A$), and  
[3.6]$-$[8.0] colours.
{\em Bottom panel.}  As top panel, but the objects are 
identified as foreground objects (open circles),  RSG (filled 
triangles), and AGB stars (open squares).
}
\label{Fig-MbolPer} 
\end{figure}

Figure~\ref{Fig-MbolPer} shows one diagnostic that was used, 
the period-luminosity diagram, with additional 
information from the pulsation amplitude and colour.

To quantify the impact of misclassifications, we compared our 
classifications to those from Jones et al.\ (2012).  Of the 
46 stars they classified as O-AGB, we classified one as SG, and 
two as C-rich.  Of the 68 stars they classified as RSGs, we 
classified seven as MA.
We also compared our classifications to those from Ruffle et 
al.\ (2015), Woods et al.\ (2011), and Jones et al.\ 
(2017b).  Of the 18 stars we classified as O-AGB, Ruffle et 
al.\ classified two as SG.  Of the 21 stars we classified as 
RSGs, they classified three as MA.
Of the 39 stars we classified as O-AGB, Woods et al.\ (2011) 
classified 0 as SG. 
Of the 13 stars we classified as RSGs, they classified zero as MA.
Of the 63 stars we classified as O-AGB, Jones et al.\ (2017b)
classified three as SG. 
Of the 57 stars we classified as RSGs, they classified one as MA. 
The differences arise mainly because we did not strictly 
enforce the classic AGB luminosity limit of $M_{\rm bol} = -7.1$~mag. 
This comparison shows that the probability of misclassification is $\sim$5-10\%.

\begin{table*}

  \caption{Carbon star sample.}
\footnotesize
  \begin{tabular}{lrrlrrrr}
  \hline
         Names                         &  R.A.      & Declination &  Identifier & Period & Amp (Filter) & Ref.\tablefootmark{a} & Remarks \\
\hline
GM 780, MACHO 213.15051.6              &   8.905250  & -73.165583  &         gm780  &  611 & 0.75 ($I$) & 1  \\
MSX SMC 029                            &   9.192958  & -73.526417  &     msxsmc029  &      &            &    \\
MSX SMC 091                            &   9.236292  & -72.421528  &     msxsmc091  &  405 & 0.57 ($K$) & 2  \\
IRAS 00350-7436                        &   9.248792  & -74.330639  &     iras00350  &      &            &     \\
MSX SMC 062, OGLE J004240.89-725705.1  &  10.670417  & -72.951583  &     msxsmc062  &  548 & 0.86 ($I$) & 1  \\
MSX SMC 054, MACHO 213.15504.265       &  10.774699  & -73.361268  &     msxsmc054  &  390 & 0.27 ($I$) & 1   \\ 
2MASS J00432649-7326433                &  10.860361  & -73.445350  &       j004326  &  327 & 0.11 ($I$) & 1  \\
MSX SMC 044, OGLE J004339.58-731457.1  &  10.914875  & -73.249333  &     msxsmc044  &  441 & 0.60 ($I$) & 1  \\
2MASS J00445256-7318258                &  11.219005  & -73.307186  &       j004452  &  158 & 0.06 ($I$) & 1  \\  
MSX SMC 105, OGLE J004502.14-725223.8  &  11.258917  & -72.873417  &     msxsmc105  &  652 & 0.93 ($I$) & 1  \\
MSX SMC 036, OGLE J004553.92-732340.7  &  11.474750  & -73.394750  &     msxsmc036  &  555 & 0.85 ($I$) & 1  \\ 
MSX SMC 014                            &  11.568042  & -74.187111  &     msxsmc014  &  361 & 1.05 ($I$) & 1   \\
MSX SMC 060, OGLE 004640.46-731646.9   &  11.668417  & -73.279778  &     msxsmc060  &  431 & 0.41 ($I$) & 1   \\ 
MSX SMC 200, smc102.5\_195             &  11.711583  & -71.794250  &     msxsmc200  &  433 & 0.36 ($I$) & 1   \\
MSX SMC 033, OGLE J004705.55-732132.5  &  11.773000  & -73.359167  &     msxsmc033  &  532 & 0.88 ($I$) & 1   \\
SSTISAGEMA J004720.02-724035.1         &  11.833320  & -72.676390  &       j004720  &  144 & 0.05 ($I$) & 1   \\
MSX SMC 66, OGLE J004852.51-730856.5   &  12.218750  & -73.149111  &     msxsmc066  &  523 & 0.46 ($I$) & 1   \\
2MASS J00485947-7335387                &  12.247937  & -73.594118  &    irasf00471  &  687 & 0.97 ($I$) & 1   \\ 
CV 78, MACHO 212.15907.1               &  12.265958  & -73.088833  &          cv78  &  435 & 0.75 ($I$) & 1   \\ 
RAW 594                                &  12.535420  & -72.838940  &        raw594  &  140 & 0.06 ($I$) & 1    \\
MSX SMC 163, MACHO 208.16031.578       &  12.753083  & -72.421806  &     msxsmc163  &  672 & 0.87 ($I$) & 1   \\
MSX SMC 142, OGLE J005140.46-725728.5  &  12.918625  & -72.958028  &     msxsmc142  &  293 & 0.76 ($I$) & 1  \\
MSX SMC 125                            &  12.958690  & -72.847080  &     msxsmc125  &  458 & 0.84 ($I$) & 1  \\ 
MSX SMC 162, OGLE J005240.16-724727.3  &  13.167375  & -72.791000  &     msxsmc162  &  535 & 0.51 ($I$) & 1  \\
MSX SMC 202, smc102.1\_11605           &  13.292208  & -72.198528  &     msxsmc202  &  486 & 0.32 ($I$) & 1  \\
MSX SMC 159, OGLE 005422.28-724329.7   &  13.592830  & -72.724940  &     msxsmc159  &      &            &    \\ 
LEGC 105, OGLE J005446.85-731337.6     &  13.695184  & -73.227182  &       legc105  &  349 & 0.62 ($I$) & 1  \\
OGLE 005450.73-730607.2                &  13.711458  & -73.102028  &      iso00548  &  430 & 0.90 ($I$) & 1  \\
OGLE 005454.09-730318.0                &  13.725417  & -73.055028  &      iso00549  &  546 & 0.56 ($I$) & 1  \\
RAW 960, OGLE J005554.61-731136.3      &  13.977620  & -73.193470  &        raw960  &  315 & 0.62 ($I$) & 1 \\
MSX SMC 209, MACHO 207.16376.687       &  14.068292  & -72.278139  &     msxsmc209  &  510 & 0.80 ($I$) & 1  \\
OGLE SMC-SC7 204803                    &  14.073042  & -72.451194  &    s3mc204803  &  194 & 0.07 ($I$) & 1  \\
IRAS 00554-7351, [GB98] S16            &  14.266458  & -73.587389  &     iras00554  &  851 & 0.83 ($I$) & 1  \\ 
MSX SMC 198, OGLE J005710.97-723059.7  &  14.295750  & -72.516639  &     msxsmc198  &  512 & 0.85 ($I$) & 1  \\
MSX SMC 155                            &  14.325625  & -72.709778  &     msxsmc155  &      &            &    \\ 
2MASS J00572054-7312460                &  14.335583  & -73.212778  &      iso00573  &  352 & 0.60 ($I$) & 1  \\
MSX SMC 093, smc107.2\_23              &  14.847333  & -73.933611  &     msxsmc093  &  463 & 0.30 ($I$) & 1  \\
OGLE 010154.53-725822.1                &  15.477417  & -72.972861  &      iso01019  &  337 & 0.51 ($I$) & 1  \\
2MASS J01045315-7204039                &  16.221485  & -72.067774  &       j010453  &      &            &    \\
MSX SMC 232, OGLE J010603.28-722231.9  &  16.513750  & -72.375611  &     msxsmc232  &  460 & 0.64 ($I$) & 1 \\
NGC 419 LE 16, smc110.2\_6965          &  17.004750  & -72.888139  &    ngc419le16  &  424 & 0.25 ($I$) & 1  \\ 
NGC 419 IR1, OGLE J010812.92-725243.7  &  17.054000  & -72.878890  &     ngc419ir1  &  456 & 0.43 ($I$) & 1 \\
NGC 419 LE 35                          &  17.072880  & -72.883720  &    ngc419le35  &  173 & 0.08 ($I$) & 1 \\
NGC 419 MIR 1                          &  17.073000  & -72.885889  &    ngc419mir1  &  738 & 0.75 ($K$) & 3  \\
NGC 419 LE 27, smc110.2\_6915          &  17.086080  & -72.881140  &    ngc419le27  &  305 & 0.08 ($I$) & 1  \\     
NGC 419 LE 18, smc110.2\_6943          &  17.103958  & -72.882472  &    ngc419le18  &  371 & 0.07 ($I$) & 1  \\
IRAS 04286-6937, MSX LMC 1007          &  67.125750  & -69.513944  &     iras04286  &  662 & 0.56 ($K$) & 4  \\
2MASS J04325737-6926331, MSX LMC 1008  &  68.239083  & -69.442528  &     iras04331  &  673 & 0.50 (W2)  & 5  \\
2MASS J04334368-7009504, MSX LMC 1077  &  68.432042  & -70.164028  &     iras04340  &  441 & 1.08 ($I$) & 1  \\
2MASS J04352409-6656493, MSX LMC 1017  &  68.850417  & -66.947028  &     iras04353  &  582 & 0.75 ($K$) & 2  \\
2MASS J04364447-7242010, MSX LMC 1067  &  69.185333  & -72.700278  &     iras04375  &  246 & 0.50 (W2) & 5 \\ %
IRAS 04374-6831, MSX LMC 1042          &  69.344667  & -68.417583  &     iras04374  &  639 & 0.71 ($K$) & 4 \\
2MASS J04425732-7012257, MSX LMC 1075  &  70.738875  & -70.207167  &     iras04433  &  486 & 0.87 ($I$) & 1 \\
2MASS J04462712-6847469                &  71.613042  & -68.796361  & sagemcj044627  &  389 & 0.91 ($I$) & 1 \\
MSX LMC 1120, lmc141.5\_9241           &  71.817042  & -68.407111  &    msxlmc1120  &  639 & 0.83 ($I$) & 1 \\
IRAS 04496-6958, MSX LMC 1130          &  72.327000  & -69.887361  &     iras04496  &  721 & 0.21 ($I$) & 1 \\ 
MSX LMC 1128                           &  72.668917  & -68.971972  &    msxlmc1128  &  442 & 0.68 ($I$) & 1 \\
\hline
\end{tabular}
\tablefoot{
\tablefoottext{a}{
1= OGLE,
2= Groenewegen et al. (2017), VMC K-band data combined with literature data,
3= Kamath et al. (2010); Amplitude estimated from their figures,
4= Whitelock et al. (2003),
5= ALLWISE + NEOWISE + SAGE + SAGE-VAR,
6= MACHO,
7= EROS,
9= redetermined combining Whitelock et al. (2003) and Wood (1998),
10= Wood, Bessell \& Paltoglou (1985),
11= Menzies  et al. (2011); Semi amplitude read-off their figures,
12= Catalina Sky Survey (Drake et al. 2014),
13= Whitelock et al. (2009),
14= Menzies et al (2010); Semi amplitude read-off their figures.
}
}

\label{Tab-Csample}
\end{table*}



\begin{table*}

  \caption{M-star sample.}
\footnotesize
  \begin{tabular}{lrrlrrrr}
  \hline
         Names                         &  R.A.      & Declination &  Identifier & Period & Amp (F) & Ref.\tablefootmark{a} & Remarks \\
\hline
WOH G 17, MSX LMC 1150                      &  69.848708  & -73.184111  &          wohg17  &  127 & 0.13 ($V$) & 12 & FG \\
MSX LMC 1212                                &  73.384667  & -69.021500  &      msxlmc1212  &      &            &    & FG \\  
RS Men, IRAS 05169-7350, MSX LMC 412        &  78.921917  & -73.787139  &           rsmen  &  304 & 0.53 ($K$) & 16 & FG \\ 
$[$W60$]$ D29, MSX LMC  819                 &  82.997208  & -66.644056  &          w60d29  &      &            &    & FG \\
HD 269788, MSX LMC  778                     &  83.723625  & -68.777639  &        hd269788  &      &            &    & FG \\ 
MSX LMC  946                                &  84.584292  & -69.625639  &       msxlmc946  &  112 & 0.14 ($I$) & 15 & FG \\
HD 269924                                   &  84.705500  & -69.451778  &        hd269924  &      &            &    & FG \\ 
MSX LMC 1677, IRAS 06013-6505               &  90.365833  & -65.089750  &      msxlmc1677  &  345 & 2.39 ($V$) & 15 & FG \\
HD 271832, MSX LMC 1687, IRAS 06045-6722    &  91.106208  & -67.388444  &        hd271832  &  541 & 0.09 ($V$) & 15 & FG \\ 
MSX LMC 1686, VV Dor,  HV 12249             &  91.699125  & -66.803472  &      msxlmc1686  &  349 & 3.01 ($V$) & 15 & FG \\
MSX SMC 067, HV 11262                       &  11.903667  & -73.078917  &       msxsmc067  &      &            &    & SG \\ 
$[$M2002$]$ SMC 10889                       &  12.112583  & -73.203417  &       smc010889  &      &            &    & SG \\ 
$[$M2002$]$ SMC 11709                       &  12.193000  & -73.472417  &       smc011709  &      &            &    & SG \\ 
PMMR24, MACHO 212.15903.1                   &  12.215904  & -73.377739  &          pmmr24  &  430 & 0.15 ($V$) & 15 & SG \\ 
MSX SMC 096                                 &  12.526667  & -73.469750  &       msxsmc096  &      &            &    & SG \\ 
MSX SMC 109                                 &  12.873625  & -73.178944  &       msxsmc109  &      &            &    & SG \\ 
MSX SMC 168, HV 1652                        &  13.861458  & -72.598931  &       msxsmc168  &      &            &    & SG \\ 
2MASS J00561387-7227324                     &  14.057833  & -72.459028  &      s3mc203963  &      &            &    & SG \\ 
2MASS J00561455-7227425                     &  14.060625  & -72.461806  &      s3mc204111  &      &            &    & SG \\ 
NGC 330 ARP 17                              &  14.079250  & -72.468972  &      s3mc205104  &      &            &    & SG \\ 
$[$M2002$]$ SMC  46662                      &  14.895872  & -72.068463  &       smc046662  &  394 & 0.12 ($V$) & 15 & SG \\ 
$[$M2002$]$ SMC  52334                      &  15.475667  & -71.871889  &       smc052334  &      &            &    & SG \\ 
PMMR 132                                    &  15.516875  & -72.436389  &         pmmr132  &      &            &    & SG \\ 
$[$M2002$]$ SMC  55188                      &  15.760263  & -72.031403  &       smc055188  &  544 & 0.32 ($I$) &  1 & SG \\  
PMMR 141                                    &  15.767833  & -72.570305  &         pmmr141  &      &            &    & SG \\ 
$[$M2002$]$ SMC  55681                      &  15.804138  & -72.157312  &        smc55681  &      &            &    & SG \\  
PMMR 145                                    &  15.814084  & -72.670111  &         pmmr145  &      &            &    & SG \\ 
HV 11464                                    &  16.039543  & -72.837717  &         hv11464  &      &            &    & SG \\ 
$[$M2002$]$ SMC  60447                      &  16.221182  & -72.796967  &  masseysmc60447  &      &            &    & SG \\ 
HV 2084                                     &  17.409256  & -73.333940  &       msxsmc149  &  737 & 0.48 ($V$) & 15 & SG \\ 
$[$M2002$]$ SMC 83593                       &  22.641504  & -73.311579  &       smc083593  &  506 & 0.47 ($V$) & 15 & SG \\ 
2MASS J04471864-6942205                     &  71.827708  & -69.705722  &   sagemcj044718  &      &            &    & SG \\
HV 2236, MSX LMC 1132                       &  72.343583  & -69.409556  &          hv2236  &  301 & 0.18 ($V$) & 18 & SG \\
HV 11423                                    &  15.229798  & -71.631369  &         hv11423  &  668 & 0.34 ($V$) & 15 & SG \\ 
GV 60, WOH S 60, iras 04537-6922            &  73.378667  & -69.297139  &            gv60  &  535 & 0.21 ($V$) & 15 & SG \\ 
MSX LMC 1189, IRAS 04553-6933               &  73.762792  & -69.486861  &      msxlmc1189  &  512 & 0.23 ($I$) & 15 & SG \\
WOH G 64, MSX LMC 1182, IRAS 04553-6825     &  73.793667  & -68.341611  &          wohg64  &  941 & 0.45 ($I$) &  1 & SG \\ 
MSX LMC 1204                                &  73.816833  & -69.320000  &      msxlmc1204  &  663 & 0.32 ($V$) & 15 & SG  \\
MSX LMC 1330                                &  73.840208  & -69.787972  &      msxlmc1330  &  646 & 0.30 ($V$) & 15 & SG \\ 
MSX LMC 1318, MACHO 17.2473.8               &  73.889750  & -69.416472  &      msxlmc1318  &  561 & 0.12 ($I$) &  1 & SG \\ 
HV 2255, MSX LMC 1328                       &  74.430458  & -70.147306  &          hv2255  &  912 & 0.17 ($K$) & 19 & SG \\
MSX LMC 1271, HD 268850                     &  75.589458  & -66.110639  &      msxlmc1271  &      &            &    & SG \\ 
HV 888, MSX LMC 43, IRAS 05042-6720         &  76.058875  & -67.270639  &           hv888  & 1005 & 0.57 ($V$) & 15 & SG \\ 
MSX LMC 141, HV 894                         &  76.389583  & -70.563028  &       msxlmc141  &  673 & 0.33 ($V$) & 15 & SG \\ 
HV 916, MSX LMC 264, IRAS 05148-6730        &  78.707167  & -67.455472  &           hv916  &  781 & 0.12 ($K$) &  4 & SG \\ 
$[$M2002$]$ LMC 116895, HV 5760             &  79.972065  & -69.459315  &       lmc116895  &      &            &    & SG \\ 
$[$M2002$]$ LMC 119219                      &  80.098395  & -69.557450  &       lmc119219  &  895 & 0.41 ($V$) & 15 & SG \\ 
WOH S 264, IRAS 05247-6941 MSX LMC 461      &  81.080458  & -69.647000  &         wohs264  & 1006 & 0.14 ($B$) &  6 & SG \\ 
$[$M2002$]$ LMC 134383, HV 957              &  81.436891  & -69.080244  &       lmc134383  &  457 & 0.24 ($V$) & 15 & SG \\ 
NGC 1948 WBT 2215, [W60] D4                 &  81.504500  & -66.271972  &  ngc1948wbt2215  &  223 & 0.34 ($V$) & 18 & SG \\ 
MSX LMC 549                                 &  81.547292  & -66.203083  &       msxlmc549  &  633 & 0.38 ($B$) &  6 & SG \\
MSX LMC 575                                 &  81.592417  & -66.357917  &       msxlmc575  &  816 & 0.06 ($V$) & 15 & SG \\ 
MSX LMC 589, $[$W60$]$ A2                   &  81.644958  & -68.861111  &       msxlmc589  &      &            &    & SG \\ 
HV 963, MSX LMC 567                         &  81.893083  & -66.891667  &           hv963  &  570 & 0.68 ($V$) & 15 & SG \\ 
HV 2551                                     &  81.910834  & -69.478917  &          hv2551  &  348 & 0.13 ($V$) & 15 & SG \\ 
\hline
\end{tabular}
\tablefoot{
\tablefoottext{a}{
1= OGLE,
2= Groenewegen et al. (2017), VMC K-band data combined with literature data,
3= Kamath et al. (2010); Amplitude estimated from their figures,
4= Whitelock et al. (2003),
5= ALLWISE + NEOWISE + SAGE + SAGE-VAR,
6= MACHO,
7= EROS,
9= period redetermined combining Whitelock et al. (2003) and Wood (1998),
10= Wood, Bessell \& Paltoglou (1985),
11= Menzies et al. (2011); Semi amplitude read-off their figures,
12= Catalina Sky Survey (Drake et al. 2014),
13= Whitelock et al. (2009),
14= Menzies et al (2010); Semi amplitude read-off their figures,
15= ASAS-3 (Pojmanski 2002),
16= Whitelock et al. (1994),
17= period redetermined by combining Catchpole \& Feast (1981), and WBF,
18= OMC (Mas-Hesse et al. 2003),
19= various literature $K$-band data.
}
}
\label{Tab-Msample}

\end{table*}

\section{Mass-loss rates, luminosities, and fits to the SEDs}
\label{App-B}

Tables~\ref{Tab-Cstar} and \ref{Tab-Mstar} list the 
parameters of the models which best fit the observed data. 
These include the identifier, information on the atmospheric 
model used (effective temperature, $\log g$, and for the C 
stars, C/O ratio), grain size and dust type, luminosity, dust 
optical depth in the $V$-band, mass-loss rate, whether $\tau$ 
was fitted (1) or fixed (0), $T_{\rm c}$, whether $T_{\rm c}$ 
was fitted or fixed, the slope of the denisity law, whether 
$p$ was fitted or fixed, outer radius (in units of inner 
radius), and the reduced $\chi^2$.
These tables are published in its entirety at the CDS; a portion is shown here for guidance regarding its form and content.

The reduced chi-square value is given for reference only.  There is a large 
range in values, and sometimes the values are far larger than unity. 
This is in part related to the construction of the SED and the 
available photometery (and error) which can be very different 
across sources, and the role of variability.

\begin{figure*}
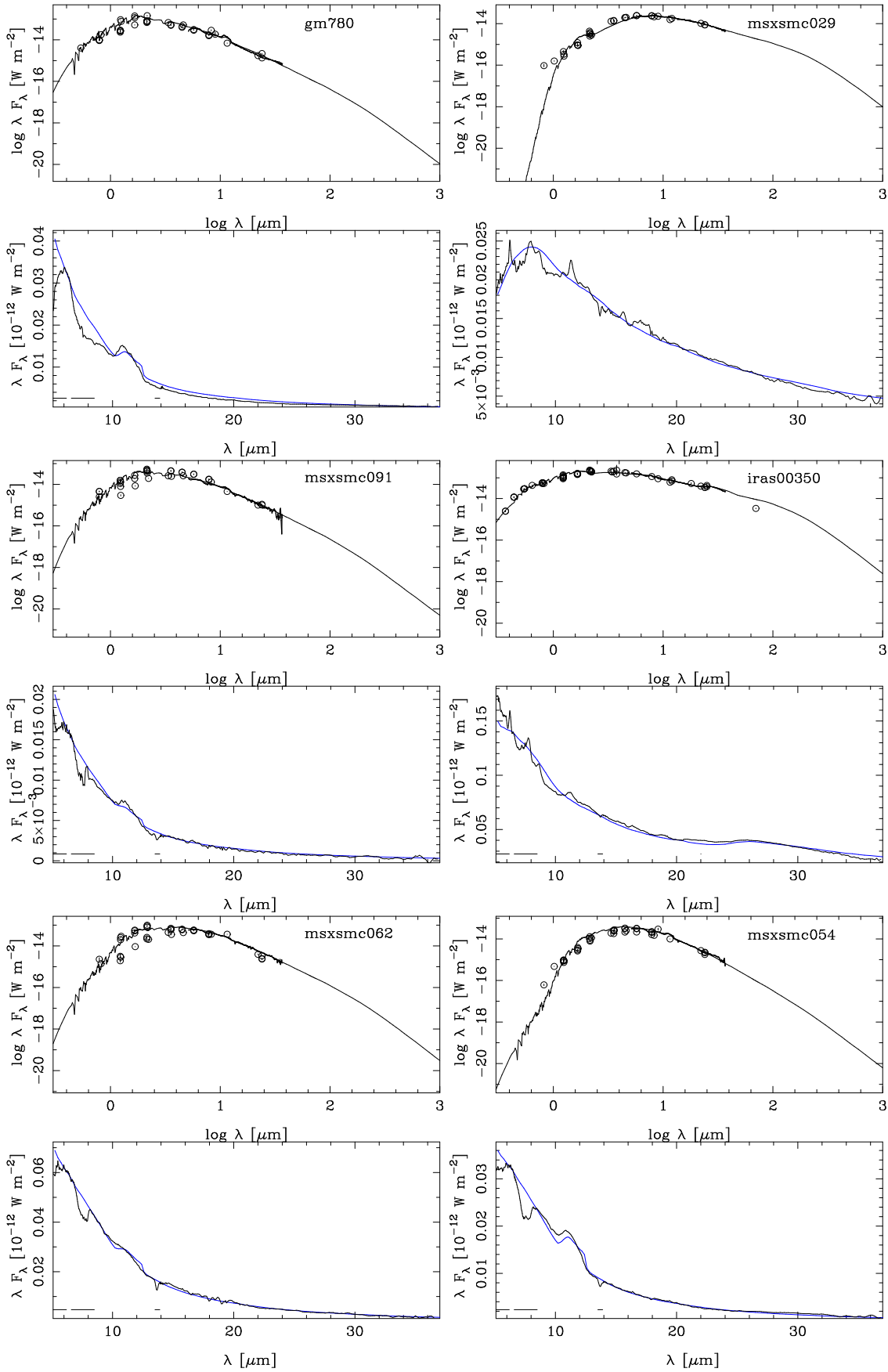

\begin{minipage}{0.40\textwidth}
\resizebox{\hsize}{!}{\includegraphics[angle=-0]{gm780_sed.ps}} 
\end{minipage}
\begin{minipage}{0.40\textwidth}
\resizebox{\hsize}{!}{\includegraphics[angle=-0]{msxsmc029_sed.ps}} 
\end{minipage}
 
\begin{minipage}{0.40\textwidth}
\resizebox{\hsize}{!}{\includegraphics[angle=-0]{msxsmc091_sed.ps}} 
\end{minipage}
\begin{minipage}{0.40\textwidth}
\resizebox{\hsize}{!}{\includegraphics[angle=-0]{iras00350_sed.ps}} 
\end{minipage}
 
\begin{minipage}{0.40\textwidth}
\resizebox{\hsize}{!}{\includegraphics[angle=-0]{msxsmc062_sed.ps}} 
\end{minipage}
\begin{minipage}{0.40\textwidth}
\resizebox{\hsize}{!}{\includegraphics[angle=-0]{msxsmc054_sed.ps}} 
\end{minipage}
 
\caption[]{ 
Fits to the SEDs and IRS spectra of C stars. 
In the upper panel the model (the continuous line) is shown with the IRS spectrum and the photometric points 
on a logarithmic scale.
The lower panel shows the IRS spectrum and model in more detail on a linear scale. 
The model has been fit to the data by scaling to a quasi-continuum point based on the 
average flux in the 6.4--6.5~\mum\ region.
Horizontal lines near the bottom indicate wavelength regions excluded from the fit.
} 

\label{Fig-Cstars} 
\end{figure*}

Figures~\ref{Fig-Cstars} and \ref{Fig-Mstars} show the 
best-fitting models, and give a subjective indication of 
the quality of the fits.  The top panel shows the observed 
SED and IRS spectrum and the fitted model on an absolute 
scale, while the bottom panel shows the IRS spectrum and the 
model (the blue line), scaled to a quasi-continuum point 
based on the average flux in the 6.35--6.55~\mum\ region.  
Horizontal lines near the bottom indicate wavelength regions 
excluded from the fit.

These figures are vavailable in the online version of the paper. 
Only some examples are shown here for guidance regarding its form and content.

\begin{table*}
\setlength{\tabcolsep}{1.3mm}

\caption{Fit results of the C-star sample.} 
\footnotesize
\begin{tabular}{rrrrrrrrrrrrr} \hline \hline 
Identifier       & $T_{\rm eff}/\log g/$(C/O) &  grain size and type   &  $L$    & $\tau_{0.5}$ & \mdot     &  f & $T_{\rm c}$ & f & $p$ & f  & $R_{\rm out}$ &  $\chi^2$   \\
                 &     (K/-/-)               &                    & (\lsol) &             & (\msolyr) &    &    (K)     &   &     &    &  ($\cdot 10^3$) &            \\
\hline 
          gm780  & 3100/+000/0140  &  a0.15 zubko100 Pitm10 Hofm0  &   20019  & 1.849e+00   &  1.09e-07  & 1 & 1200  & 0 & 2.00  & 0 & 17  &     153  \\
      msxsmc029  & 3600/+000/0140  &   a0.10 zubko100 Pitm0 Hofm0  &    4984  & 1.950e+01   &  1.26e-06  & 1 &  676  & 1 & 1.53  & 1 &  4  &      53  \\
      msxsmc091  & 2600/+000/0140  &   a0.15 zubko100 Pitm5 Hofm0  &    6975  & 3.345e+00   &  1.16e-07  & 1 & 1200  & 0 & 2.00  & 0 &  8  &     350  \\
      iras00350  & 4000/+000/0140  &  a0.15 zubko100 Pitm0 Hofm10  &   56665  & 2.585e+00   &  1.84e-07  & 1 & 1037  & 1 & 1.41  & 1 &  9.9  &      83  \\
      msxsmc062  & 2600/+000/0140  &   a0.15 zubko100 Pitm5 Hofm0  &   16243  & 5.523e+00   &  4.90e-07  & 1 &  997  & 1 & 2.00  & 0 &  8  &    1149  \\
      msxsmc054  & 2600/+000/0140  &  a0.15 zubko100 Pitm10 Hofm0  &    6632  & 1.071e+01   &  8.10e-07  & 1 & 1016  & 1 & 2.24  & 1 &  8  &     551  \\
        j004326  & 3100/+000/0140  &   a0.15 zubko100 Pitm5 Hofm0  &    4644  & 9.571e-01   &  2.50e-08  & 1 & 1200  & 0 & 2.00  & 0 & 10  &     228  \\
      msxsmc044  & 3000/+000/0140  &   a0.15 zubko100 Pitm5 Hofm0  &   16310  & 3.220e+00   &  1.76e-07  & 1 & 1200  & 0 & 2.00  & 0 &  0.037  &    1945  \\
        j004452  & 2600/+000/0140  &   a0.15 zubko100 Pitm0 Hofm0  &    9195  & 2.697e+00   &  3.21e-07  & 1 &  590  & 1 & 1.52  & 1 & 10  &      43  \\
      msxsmc105  & 2600/+000/0140  &   a0.15 zubko100 Pitm5 Hofm0  &    7900  & 6.711e+00   &  5.94e-07  & 1 &  871  & 1 & 2.00  & 0 &  8  &     181  \\
      msxsmc036  & 2600/+000/0140  &   a0.15 zubko100 Pitm5 Hofm0  &    4634  & 7.851e+00   &  4.94e-07  & 1 &  902  & 1 & 2.00  & 0 &  8  &     707  \\
      msxsmc014  & 3000/+000/0140  &   a0.15 zubko100 Pitm0 Hofm0  &    9922  & 3.608e+01   &  2.22e-06  & 1 & 1200  & 0 & 2.13  & 1 &  8  &     435  \\
      msxsmc060  & 2600/+000/0140  &   a0.15 zubko100 Pitm5 Hofm0  &   16513  & 9.044e+00   &  3.11e-07  & 1 & 1200  & 0 & 1.60  & 1 &  8  &     249  \\
      msxsmc200  & 3400/+000/0140  &   a0.15 zubko100 Pitm0 Hofm0  &    6763  & 6.000e+00   &  2.70e-07  & 1 & 1200  & 0 & 2.21  & 1 &  8  &     394  \\
      msxsmc033  & 2600/+000/0140  &   a0.15 zubko100 Pitm5 Hofm0  &   18992  & 7.168e+00   &  6.32e-07  & 1 & 1200  & 0 & 2.38  & 1 &  0.037  &     502  \\
        j004720  & 3500/+000/0140  &  a0.15 zubko100 Pitm50 Hofm0  &   16541  & 2.084e-01   &  1.45e-08  & 1 & 1200  & 0 & 2.00  & 0 & 17  &     239  \\
      msxsmc066  & 3400/+000/0140  &   a0.15 zubko100 Pitm5 Hofm0  &   17718  & 7.222e+00   &  7.87e-07  & 1 & 1200  & 0 & 2.66  & 1 &  8  &     672  \\
     irasf00471  & 3100/+000/0140  &   a0.15 zubko100 Pitm5 Hofm0  &   30455  & 8.648e+00   &  6.26e-07  & 1 & 1200  & 0 & 1.91  & 1 &  8  &     411  \\
           cv78  & 4000/+000/0140  &   a0.15 zubko100 Pitm5 Hofm0  &   10007  & 8.958e-01   &  4.00e-08  & 1 & 1200  & 0 & 2.00  & 0 & 22  &   20102  \\
         raw594  & 3300/+000/0140  &   a0.15 zubko100 Pitm5 Hofm0  &    4257  & 6.472e-01   &  1.63e-08  & 1 & 1200  & 0 & 2.00  & 0 & 10  &     208  \\
      msxsmc163  & 2600/+000/0140  &  a0.15 zubko100 Pitm10 Hofm0  &   11881  & 6.543e+00   &  4.27e-07  & 1 & 1083  & 1 & 2.00  & 0 &  8  &     132  \\
      msxsmc142  & 3300/+000/0140  &   a0.15 zubko100 Pitm5 Hofm0  &    4552  & 2.830e+00   &  7.95e-08  & 1 & 1238  & 1 & 2.00  & 0 &  0.018  &     261  \\
      msxsmc125  & 3200/+000/0140  &   a0.15 zubko100 Pitm5 Hofm0  &   14932  & 3.130e+00   &  1.61e-07  & 1 & 1200  & 0 & 2.00  & 0 &  5  &     593  \\
      msxsmc162  & 3200/+000/0140  &   a0.15 zubko100 Pitm5 Hofm0  &   13166  & 2.417e+00   &  1.14e-07  & 1 & 1200  & 0 & 2.00  & 0 &  4  &    1901  \\
      msxsmc202  & 3100/+000/0140  &   a0.15 zubko100 Pitm0 Hofm0  &   14328  & 1.968e+00   &  9.08e-08  & 1 & 1200  & 0 & 2.00  & 0 &   0.25  &     592  \\
      msxsmc159  & 2600/+000/0140  &   a0.15 zubko100 Pitm5 Hofm0  &    6512  & 1.085e+01   &  3.06e-07  & 1 & 1200  & 0 & 1.76  & 1 &  8  &     358  \\
        legc105  & 3100/+000/0140  &   a0.15 zubko100 Pitm0 Hofm0  &    7739  & 2.564e+00   &  8.84e-08  & 1 & 1200  & 0 & 2.00  & 0 & 14  &     717  \\
       iso00548  & 3400/+000/0140  &   a0.15 zubko100 Pitm5 Hofm0  &   11573  & 5.091e+00   &  3.53e-07  & 1 & 1200  & 0 & 2.37  & 1 & 12  &     237  \\
       iso00549  & 3400/+000/0140  &   a0.15 zubko100 Pitm5 Hofm0  &    9226  & 7.886e+00   &  4.49e-07  & 1 & 1105  & 1 & 2.00  & 0 &  8  &     728  \\
         raw960  & 3100/+000/0140  &   a0.15 zubko100 Pitm5 Hofm0  &    6587  & 8.716e-01   &  2.70e-08  & 1 & 1200  & 0 & 2.00  & 0 & 26  &     368  \\
      msxsmc209  & 2900/+000/0140  &   a0.15 zubko100 Pitm5 Hofm0  &   15302  & 7.203e+00   &  4.23e-07  & 1 & 1173  & 1 & 2.00  & 0 &  8  &    1380  \\
     s3mc204803  & 3200/+000/0140  &   a0.15 zubko100 Pitm5 Hofm0  &    7394  & 3.348e-01   &  1.08e-08  & 0 & 1200  & 0 & 2.00  & 0 & 13  &     163  \\
      iras00554  & 2600/+000/0140  &   a0.15 zubko100 Pitm5 Hofm0  &   23066  & 1.418e+01   &  2.56e-06  & 1 &  829  & 1 & 2.00  & 0 &  8  &    2430  \\
      msxsmc198  & 3400/+000/0140  &   a0.15 zubko100 Pitm5 Hofm0  &    7996  & 5.591e+00   &  2.08e-07  & 1 & 1249  & 1 & 2.00  & 0 &  8  &     638  \\
      msxsmc155  & 3400/+000/0140  &   a0.15 zubko100 Pitm0 Hofm0  &    8814  & 1.985e+01   &  1.09e-06  & 1 & 1200  & 0 & 2.13  & 1 &  8  &     894  \\
       iso00573  & 3200/+000/0140  &   a0.15 zubko100 Pitm5 Hofm0  &    6535  & 1.777e+00   &  5.78e-08  & 1 & 1200  & 0 & 2.00  & 0 & 15  &     289  \\
      msxsmc093  & 2600/+000/0140  &   a0.15 zubko100 Pitm5 Hofm0  &    7976  & 2.291e+00   &  8.24e-08  & 1 & 1200  & 0 & 2.00  & 0 & 23  &     340  \\
       iso01019  & 3400/+000/0140  &   a0.15 zubko100 Pitm5 Hofm0  &    6161  & 4.968e+00   &  2.84e-07  & 1 & 1200  & 0 & 2.54  & 1 & 12  &     245  \\
        j010453  & 2900/+000/0140  &   a0.15 zubko100 Pitm0 Hofm0  &    4514  & 2.951e+01   &  1.57e-06  & 1 &  997  & 1 & 2.00  & 0 &  8  &     121  \\
      msxsmc232  & 3200/+000/0140  &   a0.15 zubko100 Pitm5 Hofm0  &    6629  & 6.875e+00   &  2.55e-07  & 1 & 1200  & 0 & 2.00  & 0 &  8  &     498  \\
     ngc419le16  & 3100/+000/0140  &   a0.15 zubko100 Pitm5 Hofm0  &    5010  & 2.321e+00   &  6.66e-08  & 1 & 1200  & 0 & 2.00  & 0 & 16  &     439  \\
      ngc419ir1  & 2600/+000/0140  &   a0.15 zubko100 Pitm5 Hofm0  &   11990  & 6.016e+00   &  4.19e-07  & 1 & 1200  & 0 & 2.40  & 1 &  8  &     721  \\
     ngc419le35  & 3100/+000/0140  &   a0.15 zubko100 Pitm5 Hofm0  &    7050  & 4.350e-01   &  1.36e-08  & 0 & 1200  & 0 & 2.00  & 0 & 17  &      46  \\
     ngc419mir1  & 2600/+000/0140  &  a0.15 zubko100 Pitm5 Hofm10  &    8083  & 4.707e+01   &  2.32e-06  & 1 & 1200  & 0 & 1.87  & 1 &  8  &     257  \\
     ngc419le27  & 3200/+000/0140  &   a0.15 zubko100 Pitm5 Hofm0  &    6846  & 4.350e-01   &  1.35e-08  & 0 & 1200  & 0 & 2.00  & 0 & 17  &      70  \\
     ngc419le18  & 3200/+000/0140  &   a0.15 zubko100 Pitm5 Hofm0  &    5616  & 4.736e-01   &  1.34e-08  & 1 & 1200  & 0 & 2.00  & 0 & 19  &    1027  \\
      iras04286  & 3400/+000/0140  &   a0.15 zubko100 Pitm5 Hofm0  &   12136  & 1.327e+01   &  4.34e-07  & 1 & 1200  & 0 & 1.61  & 1 &  8  &     630  \\
      iras04331  & 3300/+000/0140  &  a0.15 zubko100 Pitm5 Hofm20  &    8380  & 2.180e+01   &  3.12e-06  & 1 &  826  & 1 & 2.00  & 0 &  8  &     256  \\
      iras04340  & 4750/+000/xxxx  &  a0.15 zubko100 Pitm10 Hofm0  &    6409  & 8.110e+00   &  5.06e-07  & 1 & 1100  & 0 & 2.01  & 1 &  0.037  &     109  \\
      iras04353  & 2600/+000/0140  &  a0.15 zubko100 Pitm5 Hofm15  &    6902  & 1.638e+01   &  1.52e-06  & 1 &  869  & 1 & 1.87  & 1 &  8  &     837  \\
      iras04375  & 2600/+000/0140  &  a0.15 zubko100 Pitm5 Hofm10  &    7257  & 1.454e+01   &  1.73e-06  & 1 &  794  & 1 & 1.96  & 1 &  8  &     269  \\
      iras04374  & 2600/+000/0140  &   a0.15 zubko100 Pitm5 Hofm0  &   12159  & 1.661e+01   &  1.39e-06  & 1 & 1004  & 1 & 2.00  & 0 &  8  &     567  \\
      iras04433  & 2600/+000/0140  &  a0.30 zubko100 Pitm10 Hofm0  &    7719  & 5.620e+00   &  4.71e-07  & 1 & 1200  & 0 & 2.12  & 1 &  8  &     244  \\
  sagemcj044627  & 2600/+000/0140  &  a0.15 zubko100 Pitm10 Hofm0  &    5325  & 2.993e+00   &  9.33e-08  & 1 & 1200  & 0 & 2.00  & 0 &  8  &     374  \\
     msxlmc1120  & 3100/+000/0140  &   a0.30 zubko100 Pitm5 Hofm0  &   15377  & 5.584e+00   &  6.07e-07  & 1 & 1006  & 1 & 1.74  & 1 &  8  &    2170  \\
      iras04496  & 2600/+000/0140  &   a0.15 zubko100 Pitm0 Hofm0  &   38738  & 4.527e+00   &  2.38e-07  & 1 & 1200  & 0 & 1.66  & 1 &   0.40  &     494  \\
\hline 
\end{tabular}
\tablefoot{
Column~1 lists the identifier.
Column~2 lists the information on the model atmosphere that is used as 
$T_{\rm eff}$ (in K) / ($\log g \cdot 100) / ($(C/O)$\cdot 100$).
Column~3 indicates the grain size and type: 
The number after the $a$ indicates the grain size in $\mu$m.
Then follows the proportion of AMC (zubko) : SiC (Pitm) : MgS (Hofm) = 100 : x : y.
Column~4 lists the luminosity.
Column~5 lists the dust optical depth at 0.5~$\mu$m, and 
Column~6 lists the total mass-loss rate (assuming an expansion velocity of 10~\ks, and a dust-to-gas ratio of 0.005).
Column~7 indicates if the optical depth was fitted (f=1, or fixed f=0).
Column~8 lists the temperature at the inner dust radius, and
Column~9 indicates if this temperature was fitted (f=1, or fixed f=0).
Column~10 lists the slope of the density law, and
Column~11 indicates if this temperature was fitted (f=1, or fixed f=0).
Column~12 lists the outer radius in units of the inner dust radius.
Column~13 lists the reduced $\chi^2$ of the fit.
}

\label{Tab-Cstar}
\end{table*}

\begin{table*} 
\setlength{\tabcolsep}{1.3mm}

\caption{Fit results of the M-star sample.} 
\footnotesize
\begin{tabular}{rrrrrrrrrrrrr} \hline \hline 
Identifier       & $T_{\rm eff}/\log g$ &  grain size and type            &  $L$     & $\tau_{0.5}$ & \mdot     &  f & $T_{\rm c}$ & f & $p$ & f  & $R_{\rm out}$ &  $\chi^2$   \\
                 &     (K/-)          &                                 & (\lsol)   &            & (\msolyr) &    &    (K)    &   &     &    & ($\cdot 10^3$)   &             \\
\hline

         wohg17  & 3300/+0.5  &         a0.10 Oliv70 AlOx30 Fe30  &  932967  & 1.201e-01   &  2.84e-07  & 1 & 1000  & 0 & 2.00  & 0 & 13  &     504  \\
     msxlmc1212  & 3900/+0.5  &         a0.20 Oliv40 AlOx60 Fe1  &  262501  & 1.000e-04   &  1.14e-10  & 0 & 1000  & 0 & 2.00  & 0 & 13  &      98  \\
          rsmen  & 2500/+0.0  &         a0.10 Oliv60 AlOx40 Fe30  &  752101  & 2.044e-01   &  4.37e-07  & 1 & 1000  & 0 & 2.00  & 0 & 13  &     492  \\
         w60d29  & 3800/+0.5  &         a0.20 Oliv90 AlOx10 Fe30  &  350174  & 1.000e-04   &  8.50e-11  & 0 & 1000  & 0 & 2.00  & 0 & 13  &     218  \\
       hd269788  & 4250/+0.0  &         a0.20 Oliv40 AlOx60 Fe3  &  611699  & 3.000e-03   &  8.67e-09  & 0 &  760  & 1 & 2.00  & 0 & 13  &    4010  \\
      msxlmc946  & 3200/+0.5  &         a0.10 Oliv50 AlOx50 Fe30  &  339561  & 4.252e-02   &  6.53e-08  & 1 & 1000  & 0 & 2.00  & 0 & 13  &      51  \\
       hd269924  & 3600/+0.0  &         a0.10 Oliv40 AlOx60 Fe3  &  745922  & 1.000e-03   &  3.69e-09  & 0 & 1000  & 0 & 2.00  & 0 & 13  &    4892  \\
     msxlmc1677  & 2700/+0.0  &         a0.10 Oliv90 AlOx10 Fe30  & 1952457  & 6.687e-01   &  2.05e-06  & 1 & 1000  & 0 & 2.00  & 0 & 13  &    1315  \\
       hd271832  & 3400/+0.0  &         a0.50 Oliv80 AlOx20 Fe3  & 1997701  & 1.391e-03   &  4.90e-09  & 1 & 1000  & 0 & 2.00  & 0 & 13  &    2553  \\
     msxlmc1686  & 2500/+0.0  &         a0.20 Oliv70 AlOx30 Fe30  & 1532676  & 2.973e-01   &  4.86e-07  & 1 & 1000  & 0 & 2.00  & 0 & 13  &     270  \\
      msxsmc067  & 3500/+0.0  &         a0.10 Oliv50 AlOx50 Fe30  &  143486  & 1.067e-02   &  1.12e-08  & 1 & 1000  & 0 & 2.00  & 0 & 13  &     196  \\
      smc010889  & 3700/+0.5  &         a0.10 Oliv70 AlOx30 Fe30  &  184822  & 5.192e-02   &  6.01e-08  & 1 & 1000  & 0 & 2.00  & 0 & 10  &     697  \\
      smc011709  & 3800/+0.5  &         a0.20 Oliv40 AlOx60 Fe30  &  114473  & 1.000e-02   &  9.16e-09  & 0 &  825  & 1 & 2.00  & 0 & 10  &      47  \\
         pmmr24  & 3800/+0.5  &         a0.50 Oliv60 AlOx40 Fe30  &   90740  & 9.816e-02   &  6.28e-08  & 1 & 1000  & 0 & 2.00  & 0 & 10  &      52  \\
      msxsmc096  & 3700/+0.5  &         a0.50 Oliv60 AlOx40 Fe30  &  107095  & 4.938e-02   &  3.35e-08  & 1 & 1000  & 0 & 2.00  & 0 & 13  &      76  \\
      msxsmc109  & 3800/+0.5  &         a0.10 Oliv90 AlOx10 Fe30  &  122980  & 1.344e-01   &  1.21e-07  & 1 & 1000  & 0 & 2.00  & 0 & 13  &     245  \\
      msxsmc168  & 3800/+0.5  &         a0.10 Oliv90 AlOx10 Fe30  &   93615  & 5.390e-02   &  4.21e-08  & 1 & 1000  & 0 & 2.00  & 0 & 13  &     320  \\
     s3mc203963  & 3900/+0.5  &         a0.20 Oliv40 AlOx60 Fe1  &   24445  & 1.289e-03   &  4.49e-10  & 0 & 1000  & 0 & 2.00  & 0 & 13  &     160  \\
     s3mc204111  & 4000/+0.0  &         a0.20 Oliv50 AlOx50 Fe30  &   57989  & 3.000e-03   &  1.25e-09  & 0 & 1000  & 0 & 2.00  & 0 & 13  &       7  \\
     s3mc205104  & 4000/+0.0  &         a0.20 Oliv40 AlOx60 Fe1  &   27422  & 1.000e-04   &  3.74e-11  & 0 & 1000  & 0 & 2.00  & 0 & 13  &      17  \\
      smc046662  & 3600/+0.5  &         a0.10 Oliv60 AlOx40 Fe30  &  113944  & 3.694e-02   &  3.42e-08  & 1 & 1000  & 0 & 2.00  & 0 & 10  &     269  \\
      smc052334  & 3700/+0.5  &         a0.10 Oliv60 AlOx40 Fe30  &   97741  & 1.352e-02   &  1.18e-08  & 1 & 1000  & 0 & 2.00  & 0 & 10  &      70  \\
        pmmr132  & 3800/+0.5  &         a0.20 Oliv70 AlOx30 Fe30  &   47998  & 9.110e-05   &  3.09e-11  & 0 & 1000  & 0 & 2.00  & 0 & 13  &     442  \\
      smc055188  & 3300/+0.5  &         a0.50 Oliv100 AlOx0 Fe30  &  104318  & 2.131e-01   &  1.38e-07  & 1 & 1000  & 0 & 2.00  & 0 & 10  &     218  \\
        pmmr141  & 3700/+0.5  &         a0.20 Oliv50 AlOx50 Fe30  &   98039  & 7.126e-03   &  3.68e-09  & 1 & 1000  & 0 & 2.00  & 0 & 13  &      82  \\
       smc55681  & 3800/+0.5  &         a0.10 Oliv70 AlOx30 Fe30  &   93022  & 6.387e-02   &  5.36e-08  & 1 & 1000  & 0 & 2.00  & 0 & 10  &     380  \\
        pmmr145  & 4000/+0.0  &         a0.20 Oliv100 AlOx0 Fe30  &   52487  & 5.000e-04   &  1.64e-10  & 0 & 1000  & 0 & 2.00  & 0 & 13  &     134  \\
        hv11464  & 3900/+0.5  &         a0.20 Oliv50 AlOx50 Fe30  &   55012  & 1.000e-02   &  7.15e-09  & 0 &  780  & 1 & 2.00  & 0 & 13  &     101  \\
 masseysmc60447  & 3800/+0.5  &         a0.50 Oliv50 AlOx50 Fe30  &   56568  & 1.770e-02   &  8.78e-09  & 1 & 1000  & 0 & 2.00  & 0 & 13  &      85  \\
      msxsmc149  & 3700/+0.5  &         a0.20 Oliv80 AlOx20 Fe30  &  198288  & 2.661e-01   &  1.76e-07  & 1 & 1000  & 0 & 2.00  & 0 & 13  &     416  \\
      smc083593  & 3600/+0.5  &         a0.10 Oliv80 AlOx20 Fe30  &  157236  & 7.268e-02   &  7.31e-08  & 1 & 1000  & 0 & 2.00  & 0 & 10  &     200  \\
  sagemcj044718  & 3600/+0.5  &         a0.10 Oliv60 AlOx40 Fe30  &   37506  & 3.201e-02   &  1.70e-08  & 1 & 1000  & 0 & 2.00  & 0 & 13  &      58  \\
         hv2236  & 3400/+0.0  &         a0.10 Oliv95 AlOx5 Fe10  &  113624  & 8.817e-01   &  8.13e-07  & 1 & 1000  & 0 & 2.00  & 0 & 13  &     716  \\
        hv11423  & 3700/+0.5  &         a0.10 Oliv80 AlOx20 Fe30  &  199246  & 7.809e-02   &  9.06e-08  & 1 & 1000  & 0 & 2.00  & 0 & 10  &     652  \\
           gv60  & 3600/+0.5  &         a0.20 Oliv80 AlOx20 Fe10  &  105755  & 5.856e-01   &  5.30e-07  & 1 &  773  & 1 & 2.00  & 0 & 13  &     115  \\
     msxlmc1189  & 3500/+0.5  &         a0.10 Oliv100 AlOx0 Fe30  &  115869  & 1.285e+00   &  1.75e-06  & 1 &  800  & 0 & 2.00  & 0 & 13  &     198  \\
         wohg64  & 3500/+0.0  &         a0.10 Oliv70 AlOx30 Fe10  &  432190  & 1.732e+01   &  3.55e-05  & 1 &  896  & 1 & 1.60  & 0 & 0.1  &   91695  \\
     msxlmc1204  & 3500/+0.0  &         a0.10 Oliv95 AlOx5 Fe30  &  198164  & 1.015e-01   &  1.85e-07  & 1 &  778  & 1 & 2.00  & 0 & 13  &     663  \\
     msxlmc1330  & 3400/+0.5  &         a0.10 Oliv80 AlOx20 Fe10  &  123820  & 6.978e-01   &  7.22e-07  & 1 & 1000  & 0 & 2.00  & 0 & 13  &     467  \\
     msxlmc1318  & 3200/+0.5  &         a0.50 Oliv70 AlOx30 Fe30  &   74948  & 3.013e-01   &  2.70e-07  & 1 &  800  & 0 & 2.00  & 0 & 13  &     163  \\
         hv2255  & 3500/+0.0  &         a0.10 Oliv90 AlOx10 Fe10  &  206364  & 6.201e-01   &  1.30e-06  & 1 &  800  & 0 & 2.00  & 0 & 13  &    1061  \\
     msxlmc1271  & 3400/+0.0  &         a0.20 Oliv70 AlOx30 Fe30  &   78591  & 5.075e-01   &  2.02e-07  & 1 & 1000  & 0 & 2.00  & 0 & 13  &      95  \\
          hv888  & 3500/+0.5  &         a0.50 Oliv70 AlOx30 Fe3   &  526751  & 2.736e-01   &  4.76e-07  & 1 & 1044  & 1 & 2.00  & 0 & 13  &   13598  \\
      msxlmc141  & 3600/+0.5  &         a0.10 Oliv80 AlOx20 Fe10  &  135417  & 5.272e-01   &  6.10e-07  & 1 & 1000  & 0 & 2.00  & 0 & 13  &     738  \\
          hv916  & 3400/+0.5  &         a0.20 Oliv90 AlOx10 Fe30  &  104646  & 1.555e+00   &  6.95e-07  & 1 & 1000  & 0 & 2.00  & 0 & 13  &     787  \\
      lmc116895  & 3800/+0.5  &         a0.20 Oliv80 AlOx20 Fe30  &  124490  & 1.014e-01   &  1.01e-07  & 1 &  758  & 1 & 2.00  & 0 & 10  &     138  \\
      lmc119219  & 3900/+0.5  &         a0.20 Oliv95 AlOx5 Fe30   &   55661  & 1.100e+00   &  4.04e-07  & 1 & 1000  & 0 & 2.00  & 0 & 10  &     363  \\
        wohs264  & 3400/+0.5  &         a0.20 Oliv90 AlOx10 Fe30  &  158954  & 1.249e+00   &  8.71e-07  & 1 &  912  & 1 & 2.00  & 0 & 13  &     161  \\
      lmc134383  & 3500/+0.5  &         a0.10 Oliv90 AlOx10 Fe30  &   87053  & 2.377e-01   &  2.79e-07  & 1 &  800  & 0 & 2.00  & 0 & 10  &     349  \\
 ngc1948wbt2215  & 3500/+0.0  &         a0.20 Oliv40 AlOx60 Fe30  &  227669  & 1.611e-02   &  2.30e-08  & 1 &  830  & 1 & 2.00  & 0 & 13  &      79  \\
      msxlmc549  & 3200/+0.5  &         a0.10 Oliv90 AlOx10 Fe30  &  178966  & 3.693e-01   &  3.51e-07  & 1 & 1000  & 0 & 2.00  & 0 & 13  &     565  \\
      msxlmc575  & 3800/+0.5  &         a0.20 Oliv70 AlOx30 Fe30  &  163654  & 4.616e-02   &  3.78e-08  & 1 &  893  & 1 & 2.00  & 0 & 13  &      29  \\
      msxlmc589  & 3900/+0.5  &         a0.50 Oliv70 AlOx30 Fe30  &  230016  & 2.711e-01   &  2.88e-07  & 1 & 1000  & 0 & 2.00  & 0 & 13  &    1933  \\
          hv963  & 3700/+0.5  &         a0.50 Oliv80 AlOx20 Fe30  &  104486  & 2.945e-01   &  3.91e-07  & 1 &  758  & 1 & 2.00  & 0 & 13  &     107  \\
         hv2551  & 3600/+0.5  &         a0.10 Oliv80 AlOx20 Fe30  &   87885  & 6.232e-02   &  4.67e-08  & 1 & 1000  & 0 & 2.00  & 0 & 13  &     168  \\
         hv2561  & 3700/+0.5  &         a0.20 Oliv90 AlOx10 Fe10  &  216781  & 4.956e-01   &  6.68e-07  & 1 &  758  & 1 & 2.00  & 0 & 13  &     146  \\
         hv5870  & 3600/+0.5  &         a0.10 Oliv90 AlOx10 Fe10  &   72629  & 1.054e+00   &  8.76e-07  & 1 & 1000  & 0 & 2.00  & 0 & 13  &     449  \\
\hline 
\end{tabular} 
\tablefoot{
Column~1 lists the identifier.
Column~2 lists the information on the model atmosphere that is used as 
$T_{\rm eff}$ (in K) / $\log g$.
Column~3 indicates the grain size and type: 
The number after the $a$ indicates the grain size in $\mu$m.
Then follows the proportion of olivine (MgFeSiO$_4$), corundum, metallic iron, and sometimes forsterite.
Column~4 lists the luminosity.
Column~5 lists the dust optical depth at 0.5~$\mu$m, and 
Column~6 lists the total mass-loss rate (assuming an expansion velocity of 10~\ks, and a dust-to-gas ratio of 0.005).
Column~7 indicates if the optical depth was fitted (f=1, or fixed f=0).
Column~8 lists the temperature at the inner dust radius, and
Column~9 indicates if this temperature was fitted (f=1, or fixed f=0).
Column~10 lists the slope of the density law, and
Column~11 indicates if this temperature was fitted (f=1, or fixed f=0).
Column~12 lists the outer radius in units of the inner dust radius.
Column~13 lists the reduced $\chi^2$ of the fit.
}
\label{Tab-Mstar}
\end{table*}

\clearpage

\clearpage

\setcounter{figure}{1}

\begin{figure*}
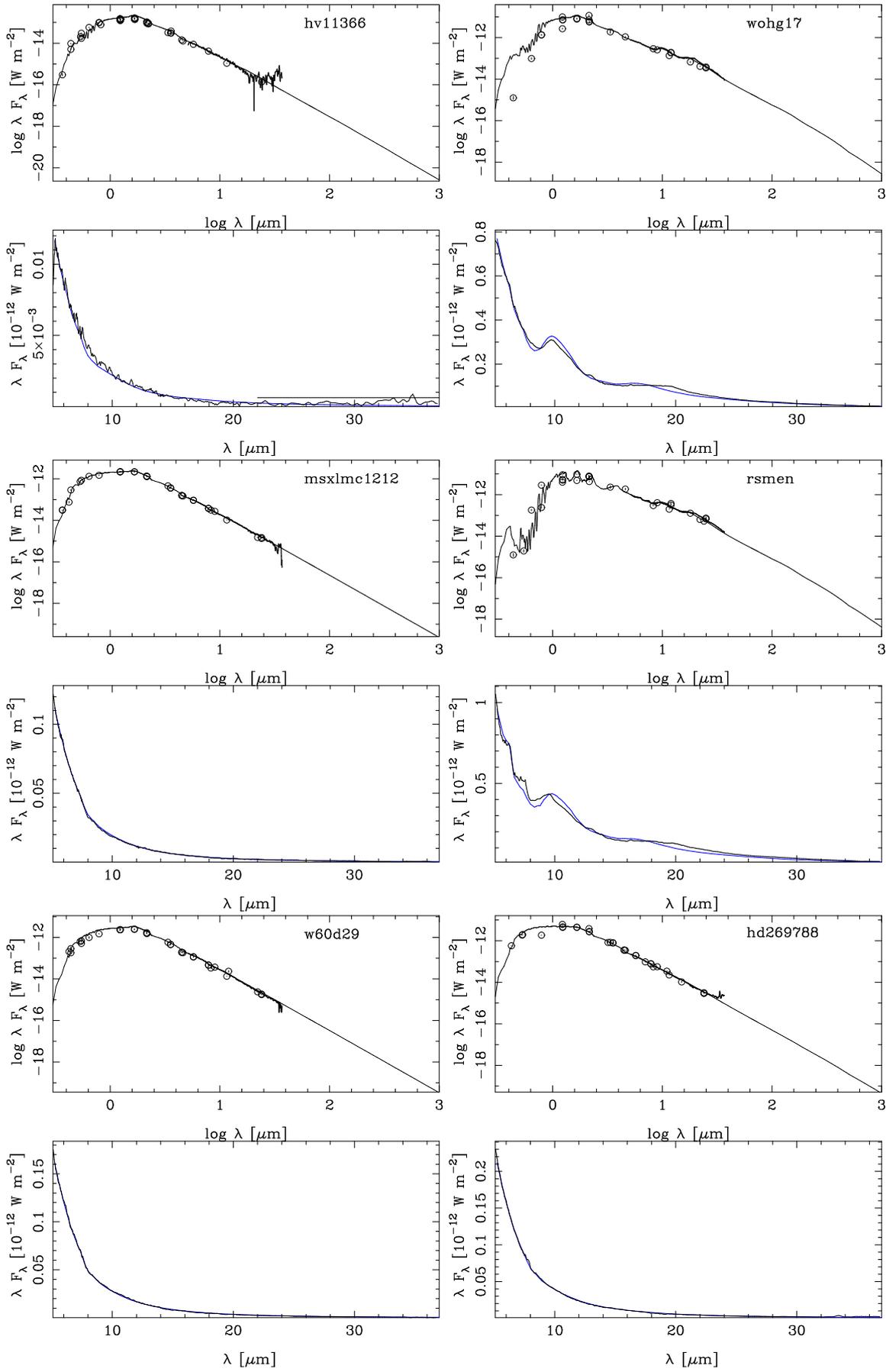

\begin{minipage}{0.41\textwidth}
\resizebox{\hsize}{!}{\includegraphics[angle=-0]{hv11366_sed.ps}} 
\end{minipage}
\begin{minipage}{0.41\textwidth}
\resizebox{\hsize}{!}{\includegraphics[angle=-0]{wohg17_sed.ps}} 
\end{minipage}
 
\begin{minipage}{0.41\textwidth}
\resizebox{\hsize}{!}{\includegraphics[angle=-0]{msxlmc1212_sed.ps}} 
\end{minipage}
\begin{minipage}{0.41\textwidth}
\resizebox{\hsize}{!}{\includegraphics[angle=-0]{rsmen_sed.ps}} 
\end{minipage}
 
\begin{minipage}{0.41\textwidth}
\resizebox{\hsize}{!}{\includegraphics[angle=-0]{w60d29_sed.ps}} 
\end{minipage}
\begin{minipage}{0.41\textwidth}
\resizebox{\hsize}{!}{\includegraphics[angle=-0]{hd269788_sed.ps}} 
\end{minipage}
 
\caption[]{As Figure~\ref{Fig-Cstars} for the O stars.} 
\label{Fig-Mstars} 
\end{figure*}

\clearpage

\section{Synthetic magnitudes}
\label{App-C}

Based on the best-fitting model, magnitudes and fluxes 
are computed in a large number of filters for every star. 
The procedure follows that in Groenewegen (2006).
The calibration is based on the Vega model of Bohlin (2007) 
and assumes zero magnitudes in all filters.

The filters that are provided are: 
Bessell $UBVRI$, 2MASS $JHK$, VISTA $ZYJHK$, WISE 1-4,
IRAC $[3.6],[4.5],[5.8],[8.0]$, MIPS $[24], [70], [160]$,
Herschel PACS $[70], [110], [170]$, SPIRE $[250], [350], 
[550]$, Akari N2, N3, N4, S7, S9W, S11, L15, L18W, L24, N60, 
WS, WL, N160, and generic sub-mm filters at 350, 450, 850 
and 1300 $\mu$m.

Of interest may be the magnitudes in a large number of wide 
and medium band filters
from
NIRCAM\footnote{http://www.stsci.edu/jwst/instruments/nircam/instrumentdesign/filters/}
(F070W,  F090W, F115W, F150W, F200W, F277W, F356W,  F444W,
F140M,  F162M, F182M, F210M, F250M, F300M, F335M, F360M, 
F410M, F430M, F460M, F480M) and
MIRI\footnote{Alistair Glasse, private communication, December 2015.} (F560W, F770W, F1000W, F1130W, F1280W, 
F1500W, F1800W, F2100W, F2550W) instruments.

\begin{figure}
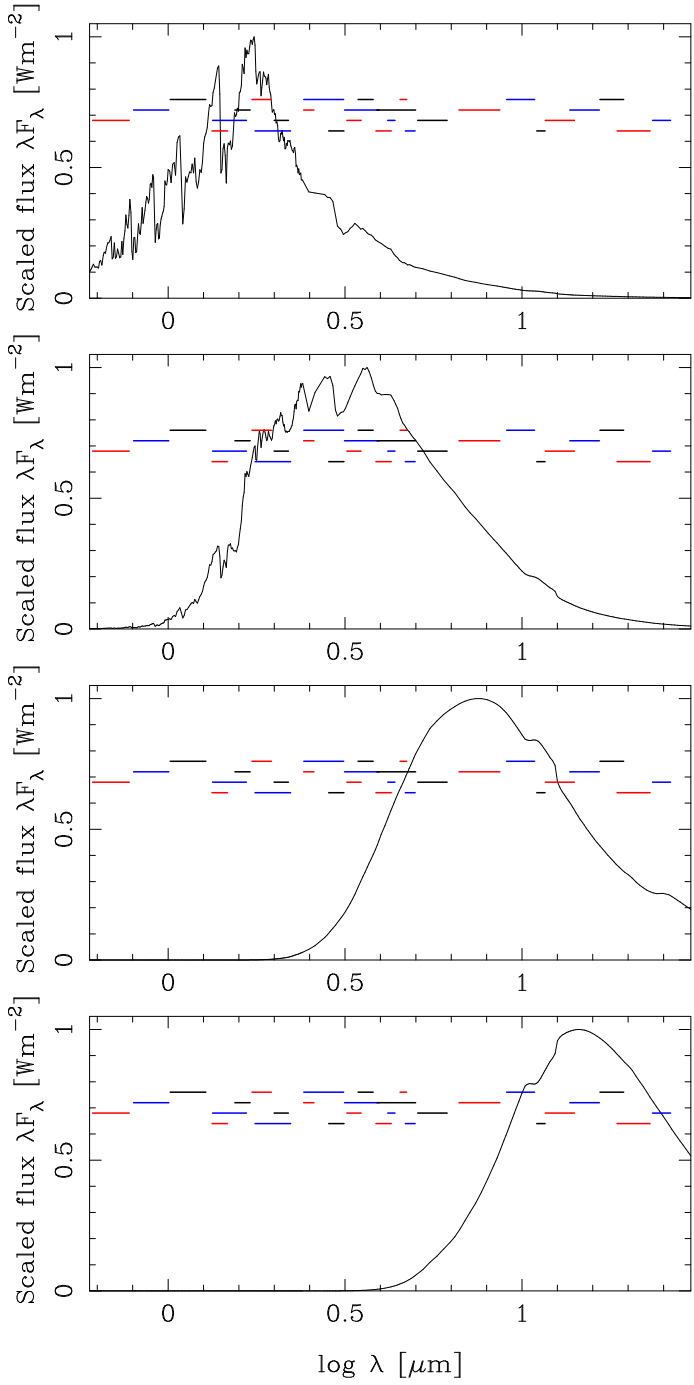


\begin{minipage}{0.49\textwidth}
\resizebox{\hsize}{!}{\includegraphics{SedFilter_ngc419le18.ps}}
\end{minipage}

\begin{minipage}{0.49\textwidth}
\resizebox{\hsize}{!}{\includegraphics{SedFilter_ngc419ir1.ps}}
\end{minipage}

\begin{minipage}{0.49\textwidth}
\resizebox{\hsize}{!}{\includegraphics{SedFilter_ngc419mir1.ps}}
\end{minipage}

\begin{minipage}{0.49\textwidth}
\resizebox{\hsize}{!}{\includegraphics{SedFilter_ERO0518484.ps}}
\end{minipage}

\caption[]{Normalized SEDs of four C stars with little, 
moderate, strong, and extreme mass-loss rates (NGC 419 LE18, 
IR1, MIR1, and ERO0518484 from top to bottom).
The horizontal bars indicate the FWHM range of 29 NIRCAM and 
MIRI filters.
}
\label{Fig-JWST-C} 
\end{figure}

\begin{figure}
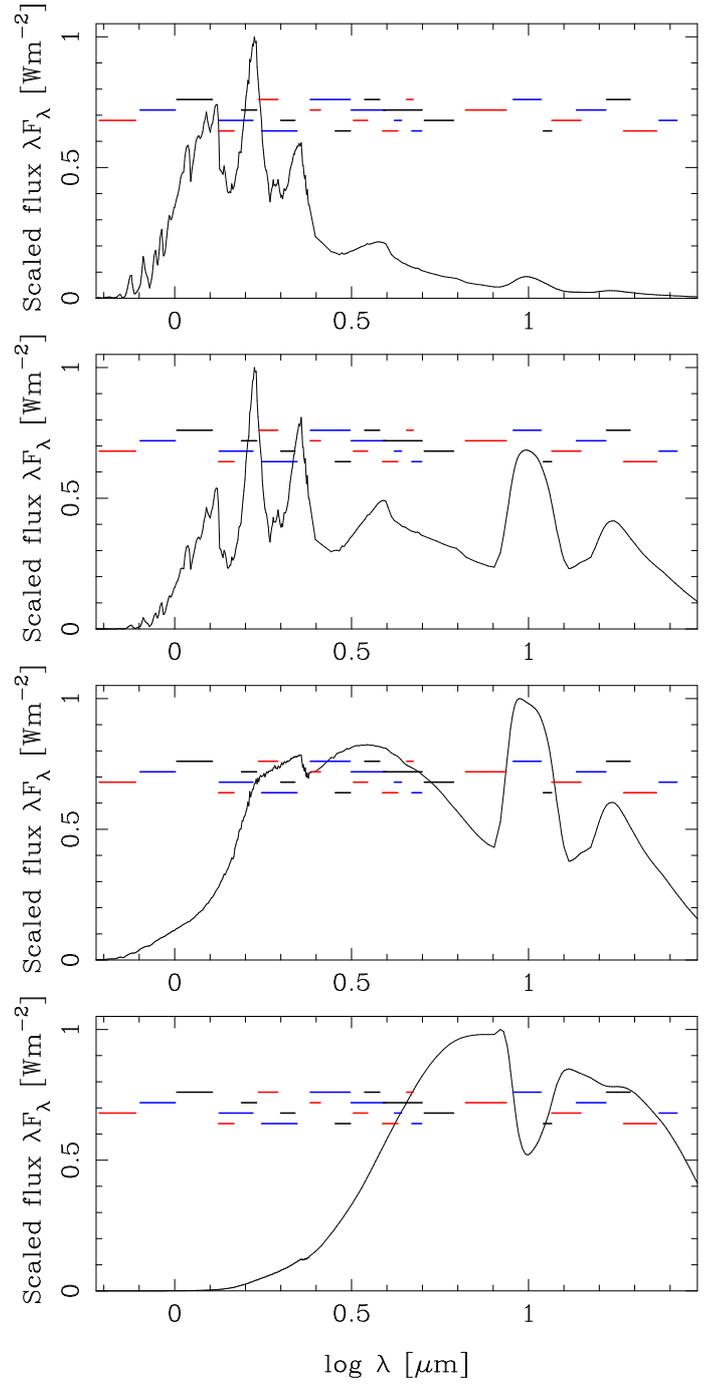


\begin{minipage}{0.49\textwidth}
\resizebox{\hsize}{!}{\includegraphics{SedFilter_msxlmc1677.ps}}
\end{minipage}

\begin{minipage}{0.49\textwidth}
\resizebox{\hsize}{!}{\includegraphics{SedFilter_w60a72.ps}}
\end{minipage}

\begin{minipage}{0.49\textwidth}
\resizebox{\hsize}{!}{\includegraphics{SedFilter_iras05003.ps}}
\end{minipage}

\begin{minipage}{0.49\textwidth}
\resizebox{\hsize}{!}{\includegraphics{SedFilter_iras05246.ps}}
\end{minipage}

\caption[]{Normalized SEDs of four O stars with little, 
moderate, strong, and heavy mass-loss rates (MSX LMC 1677, 
W60 A72, IRAS 05003, and IRAS 05246 from top to bottom).
The horizontal bars indicate the FWHM range of 29 NIRCAM and 
MIRI filters.
}
\label{Fig-JWST-O} 
\end{figure}

Figures~\ref{Fig-JWST-C} and \ref{Fig-JWST-O} show the 
normalized SEDs of four C and O stars, respectively, with
the NIRCAM and MIRI filters indicated by the FWHM of the 
response curves. 

Table~\ref{Tab-Synphot}, available at the CDS, shows an example 
of the synthetic photometry for one star.  In the electronic version 
all filters are listed on one single row per object.
All values are in magnitudes, except the three PACS, three SPIRE and 
four sub-mm filters which are given in $\log F$(mJy).  The 
magnitudes and fluxes can be scaled using the adopted 
distances (see Section~\ref{themodel}) and 
fitted luminosities (Tables~\ref{Tab-Cstar} and 
\ref{Tab-Mstar}) for each star.

\begin{sidewaystable}
\vspace{67mm}
\setlength{\tabcolsep}{1.58mm}
\caption{Synthetic photometry. Example for one object.}
\centering
\begin{tabular}{cccccccccccccccccccc} 
\hline\hline             
\multicolumn{5}{c}{Bessell} & \multicolumn{3}{c}{2MASS} & \multicolumn{5}{c}{VISTA} & \multicolumn{4}{c}{WISE}  \\  
    U  & B & V & R & I      &       J &  H & K          &   Z & Y & J & H & K        & W1 & W2 & W3 & W4  \\ \hline
22.332 & 20.869 & 18.654 & 17.109 & 15.629 & 12.970 & 11.698 & 11.192 & 15.057 & 14.075 & 12.900 & 11.726 & 11.186 & 10.508 & 9.314  & 6.516  & 5.217 \\ \hline\hline
\multicolumn{4}{c}{IRAC}         & \multicolumn{3}{c}{MIPS} & \multicolumn{3}{c}{PACS} & \multicolumn{3}{c}{SPIRE} & \multicolumn{4}{c}{Akari}   \\  
$[3.6]$ & [4.5] & [5.8] & [8.0]  &   [24]  & [70] & [160]   &         70 & 110 & 170   &        250 & 350 & 550    & N2 & N3 & N4 & S7 \\ \hline
10.298 & 9.419  & 8.598 & 7.477 & 5.123 & 3.374 & 2.150 & 1.551 & 1.477 & 1.327 & 1.096 & 0.816 & 0.431 & 11.075 & 10.654 & 9.502 & 7.760 \\ \hline\hline
\multicolumn{9}{c}{Akari}                           & \multicolumn{8}{c}{JWST}   \\  
S9 & S11 & L15 & L18W & L24 & N60 & WS & WL & N160 & F070W & F090W & F115W & F150W & F200W & F277W & F356W & F444W  \\ \hline
7.206 & 6.731 & 5.905 & 5.591 & 5.148 & 3.519 & 3.107 &  2.250 & 2.093 & 16.604 & 14.969 & 13.479 & 12.190 & 11.210 & 11.023 & 10.333 & 9.486 \\ \hline \hline
\multicolumn{17}{c}{JWST}   \\  
F770W & F1000W & F1130W & F1280W & F1500W & F1800W & F2100W & F2550W & F140M & F158M & F162M & F182M & F210M & F250M & F380M & F300M & F335M \\ \hline
7.555 & 6.846 & 6.572 & 6.303 & 5.970 & 5.630 & 5.347 & 4.996 & 12.384 & 11.832 & 11.808 & 11.241 & 11.172 & 11.100 & 9.982 & 11.147 & 10.622 \\ \hline \hline
\multicolumn{6}{c}{JWST}                        &  \multicolumn{4}{c}{sub-mm}     & identifier   \\
F360M & F410M & F430M & F460M & F480M & F560W   &   350 &   450 &   850  &   1300 &               \\ \hline
10.101 & 9.762 & 9.519 & 9.351 & 9.187 & 8.619  & 0.820 & 0.589 & -0.172 & -0.666 &      j004452  \\
\hline
\hline
\end{tabular}

\label{Tab-Synphot}
\end{sidewaystable}

\end{appendix}

\end{document}